\documentclass[fleqn,usenatbib]{mnras}
\usepackage{pdflscape}
\usepackage{float,lscape}
\usepackage{adjustbox}

\usepackage[T1]{fontenc}
\usepackage{ae,aecompl}
\usepackage{longtable}
\usepackage[table, dvipsnames]{xcolor}
\usepackage{comment}

\usepackage{newtxtext,newtxmath}
\newcommand{\sqdeg}{\mbox{$\mbox{deg}^2$}}

\usepackage[graphicx]{realboxes}
\usepackage{graphicx}	

\usepackage{tikz}
\usepackage{pgfplots}
\pgfplotsset{compat=1.14}
 \DeclareGraphicsExtensions{.png,.pdf}

\title[DDOTI Observations During O3]{DDOTI Observations of Gravitational-Wave Sources Discovered in O3}

\author[R.~L.~Becerra et al.]{%
R.~L.~Becerra,$^{1}$\thanks{E-mail: rosa.becerra@correo.nucleares.unam.mx (RLB)}
S.~Dichiara,$^{2,3}$
A.~M.~Watson,$^{4}$
E.~Troja$^{2,3}$ 
N.~R.~Butler,$^{5}$\newauthor
M.~Pereyra,$^{6}$
E.~Moreno~M\'endez,$^{7}$
F.~De Colle,$^{1}$
W.~H.~Lee,$^{4}$
A.~S.~Kutyrev$^{2,3}$ \newauthor
and 
K.~O.~C.~L\'opez$^{4}$
\\
$^1$ Instituto de Ciencias Nucleares, Universidad Nacional Aut\'onoma de M\'exico, Apartado Postal 70-264, 04510 M\'exico, CDMX, Mexico\\
$^2$ Department of Astronomy, University of Maryland, College Park, MD 20742-4111, USA\\
$^3$ Astrophysics Science Division, NASA Goddard Space Flight Center, 8800 Greenbelt Road, Greenbelt, MD 20771, USA\\
$^4$ Instituto de Astronom{\'\i}a, Universidad Nacional Aut\'onoma de M\'exico, Apartado Postal 70-264, 04510 M\'exico, CDMX, Mexico\\
$^5$ School of Earth and Space Exploration, Arizona State University, Tempe, AZ 85287, USA\\
$^6$ CONACYT, Instituto de Astronom{\'\i}a, Universidad Nacional Aut\'onoma de M\'exico, 22860 Ensenada, BC, Mexico\\
$^7$ Facultad de Ciencias, Universidad Nacional Aut\'onoma de M\'exico, A. P. 70-543, 04510 D.F, Mexico
}

\pubyear{2020}

\begin{document}
\label{firstpage}
\pagerange{\pageref{firstpage}--\pageref{lastpage}}
\maketitle

\begin{abstract}

We present optical follow-up observations with the DDOTI telescope of gravitational-wave events detected during the Advanced LIGO and Advanced Virgo O3 observing run. 
DDOTI is capable of responding to an alert in a few minutes, has an instantaneous field of about 69 deg$^{2}$, and obtains $10\sigma$ upper limits of $w_{\rm lim}=18.5$ to 20.5 AB mag in 1000~s of exposure, depending on the conditions. 
We observed 54\% (26 out of 48) of the unretracted gravitational-wave alerts and did not find any electromagnetic counterparts.
We compare our upper limits to various possible counterparts: the kilonova AT~2017gfo, models of radioactive- and magnetar-powered kilonovae, short gamma-ray burst afterglows, and AGN flares. 
Although the large positional uncertainties of GW sources do not allow us to place strong constraints during O3, DDOTI observations of well-localized GW events in O4 and beyond could meaningfully constrain models of compact binary mergers.
We show that DDOTI is able to detect kilonovae similar to AT~2017gfo up to about 200~Mpc and magnetar-powered kilonovae up to 1~Gpc.
We calculate that nearby ($\lesssim$200 Mpc) afterglows have a high chance ($\approx$70\%) to be detected by rapid ($\lesssim$3 hours) DDOTI observations if observed on-axis, whereas off-axis afterglows are unlikely to be seen.
Finally, we suggest that long-term monitoring of massive BBH events with DDOTI could confirm or rule out late AGN flares associated with these events.

\end{abstract}


\section{INTRODUCTION}
\label{sec:introduction}

The direct detection of gravitational waves (GWs) 
has opened a new window to a hitherto invisible universe and allows us to exploit the long-promised advantages of multi-messenger astronomy \citep{LVC15,Abbott2016b,Abbott2017a,Abbott2017b,Abbott2019}.

At first, the Advanced LIGO observatories detected GW signals from the mergers of binary black holes (BBH), starting with GW150914 \citep{Abbott2016a}.
Although these systems are generally not expected to produce bright electromagnetic (EM) signals, 
many facilities searched for EM counterparts  \citep[e.g.,][]{Connaughton2016,Evans2016,Savchenko2016,Troja2016,Golkhou2018}. 

Subsequently, the Advanced LIGO and Virgo observatories 
discovered and localized GW170817, 
a GW signal from the inspiral of two low-mass compact objects consistent with a binary neutron star (BNS) merger \citep{Abbott2017a}. 
The simultaneous detection of a short gamma-ray burst (sGRB) by the {\itshape Fermi}/GBM and {\itshape INTEGRAL}/SPIACS gamma-ray detectors \citep{Goldstein2017,Savchenko2017} established the association between BNS mergers and at least some sGRBs. 
This was followed by the discovery of the bright kilonova AT~2017gfo
\citep{Coulter2017,Arcavi2017,Diaz2017,Evans2017,Lipunov2017,Pian2017,Soares2017,Tanvir2017,Troja2017,Valenti2017} and the off-axis GRB afterglow 
\citep{Troja2017,Hallinan2017,Lazzati2018}. 

With these antecedents, the third run of 
the Advanced LIGO and Virgo observatories (hereafter referred to as O3) began on 2019 April 1. 
The improved sensitivity of the GW detectors increased the rate of events, but also their average distance,
so that the EM counterpart of an event similar to GW170817 might easily escape detection.  
However, studies of sGRBs suggest that NS mergers could lead to a wide variety of EM counterparts, some of which could be detectable at 200 Mpc and beyond \citep[e.g.][]{Sagues2021,Nakar2020,Ascenzi2021}. 
For instance, X-rays observations of sGRB afterglows have been used to argue that a large ($\gtrsim$60\%) 
fraction of BNS mergers could produce a long-lived magnetar remnant \citep[e.g.][]{Gao2006,Rowlinson2013,Lu2015}, although 
this number was recently revised to $<$40\% based on late-time radio monitoring of nearby sGRBs \citep{Ricci2021,Bruni2021}. 
A long-lived NS remnant could substantially affect the properties of the EM counterpart, producing optical transients up to a hundred times brighter than radioactively powered kilonovae \citep[e.g., ][]{2013Yu, Gao2017}. These magnetar-powered kilonovae could be visible with telescopes of modest apertures up to $z\sim0.2$ \citep{Yuan2021}.

BNS mergers are not the only possible GW sources potentially capable of emitting bright EM radiation. 
Mergers between neutron stars and black holes (NSBHs) are also promising candidates for the production
of sGRBs and kilonovae, depending on the NS equation of state (EoS), the mass ratio, the BH spin, and the orbital characteristics of the binary merger \citep{Kawaguchi2015,Kruger2020,Fernandez2020}. 
In favorable conditions, that is a BH with high spin ($\chi>0.8$) and low mass ($<10M_{\odot}$), a NSBH merger could power a kilonova brighter than AT~2017gfo \citep{Barbieri2019,Barbieri2020}. 
BBH mergers have also been suggested as sources of detectable EM emission \citep[e.g., ][]{Perna2016,Zhang2016,McKernan2020}.

Unfortunately, in spite of the high number of GW alerts and the efforts of several observing groups, no EM counterpart was unambiguously identified during O3, either by us or by others
\citep{Antier2020,Gompertz2020,Page2020,Klinger2021,Skymapper}. 
Despite this, these observations and the observations we report here place interesting upper limits on the properties of the EM counterparts of several GW sources. 
For example, the event GW190814,
produced by the merger of a massive BH and a lighter object \citep{LVC190814} 
was extensively observed  
\citep{Dobie2019,Watson2020,Ackley2020,Andreoni2020,Vieira2020,Thakur2020,Morgan2020,Alexander2021}.
\cite{Watson2020} and \cite{Thakur2020} excluded the possibility of an on-axis sGRB, \cite{Alexander2021} disfavored an sGRB seen off-axis, \cite{Thakur2020}, \cite{Andreoni2020} and \cite{Ackley2020} used the optical and infrared observations to constrain the ejecta mass. These studies stimulated the development of new and more sophisticated kilonova models, taking into account different progenitor masses, viewing angles, and BH spins \citep[e.g.][]{Bernuzzi2020,Kawaguchi2020,Darbha2020,Korobkin2021,Wollaeger2021} and also helped fine-tune the observing strategy of ground-based telescopes responding to LIGO and Virgo alerts. 

During O3, GW candidate sources were localized over very large areas of the sky ($\gtrsim$1000 deg$^2$)
and at large distances ($\gtrsim$200~Mpc), making galaxy-targeted strategies much less effective \citep{Gehrels2016}. 
This represents a challenge for telescopes with narrow field of views and
requires the use of wide-field instruments, such as the Deca-Degree Optical Transient Imager (DDOTI),
to search the GW area and identify the best candidates for further, deeper follow-up. 
DDOTI \citep{Watson2016} was designed with the purpose of catching optical counterparts from poorly localized GRBs, neutrino events, and GW sources. The main feature of DDOTI is its wide instantaneous field of view of about 69 deg$^2$. Moreover, DDOTI has an AB $10\sigma$ limiting magnitude of $w_{\rm lim}=18.5-20.5$ for 1000 seconds of exposure, depending on the conditions \citep{Watson2020,Thakur2020}.

In this paper, we summarise the DDOTI follow-up campaign of GW alerts during the LIGO/Virgo O3 run. In \S\ref{sec:o3}, we briefly describe the O3 run of the Advanced LIGO and Advanced Virgo detectors. In \S\ref{sec:ddoti} we present our observing campaign with the DDOTI telescope. 
In \S\ref{sec:results} we outline different constrains derived from our upper limits on the optical emission of the GW sources and make predictions for O4. Finally, in \S\ref{sec:summary} we summarise our results and discuss their implications. 

\section{Advanced LIGO/Virgo Observations}
\label{sec:o3}

The O3 run of Advanced LIGO and Advanced Virgo lasted from 2019 April 1 15:00 UTC to 2020 March 27 17:00 UTC, ending earlier than planned in response to the COVID-19 health emergency. The run was split into “O3a” and “O3b” by a month-long commissioning break in 2019 October. In total, the run consisted of 330 days.

The LIGO Scientific Collaboration and the VIRGO Collaboration (LVC) sent real-time alerts of events to the astronomical community through the Gamma-Ray Coordinates Network (GCN) and published their preliminary analysis on their  Gravitational-Wave Candidate Event Database (GraceDB) website (\url{https://gracedb.ligo.org/superevents/public/O3/}). The analysis includes classification and probability maps, produced by the {\sc Bayestar} \citep{Singer2016} and {\sc LALInference} \citep{Veitch2015} pipelines. The localizations and classifications were often improved over the first few hours, and these improvements were communicated by further notices and updates on the website.


During O3 there were 80 automated preliminary GCN Notices, 56 of them were not retracted. Furthermore, 3 of unretracted events have a probability of being terrestrial larger than 50\%. Discarding these, we are left with with 31 and 22 candidate events discovered during O3a and O3b respectively.
The final analysis and classification of the events
detected during O3a are reported in the Gravitational Wave Transient Catalog 2 \citep[][hereafter referred to as GWTC-2]{Abbott2020c}.
Out of the 31 candidates reported in O3a, 5 had a signal to noise ratio above threshold in only one detector \citep{Abbott2020c}.
In our analysis here, we include only the 26 events reported in the GWTC-2 catalog whose signal was triggered in more than one detector during O3a and all 22 non-retracted events detected during O3b. That is, we consider a total of 48 events.

Among the confirmed events in O3a,  GW190425 is consistent with a BNS merger, GW190426\_152155 is a candidate NSBH merger,
and GW190814 is classified as a mass gap (MG) event,
that is a compact binary system with at least one object with mass in the range between known NSs and BHs  ($2.5\,M_\odot < M < 5\,M_\odot$). All the other GW sources detected during O3a are consistent with BBH mergers.

For events in O3b, we rely on the preliminary analysis reported in GraceDB. Our results for events in O3b should be tempered by the knowledge that some of these events might be withdrawn when the final analysis is published.
Among the currently unretracted events in O3b, S191205ah is consistent with a NSBH merger, S191213g and S200213t with BNS mergers, S200115j and S200316bj with MG events. 
S200114f with a burst which is defined as a signal that is detected without a template and without prior knowledge of the waveform. 
The remaining sources are consistent with BBH mergers.

\begin{figure}
\centering
\includegraphics[width=0.45\textwidth]{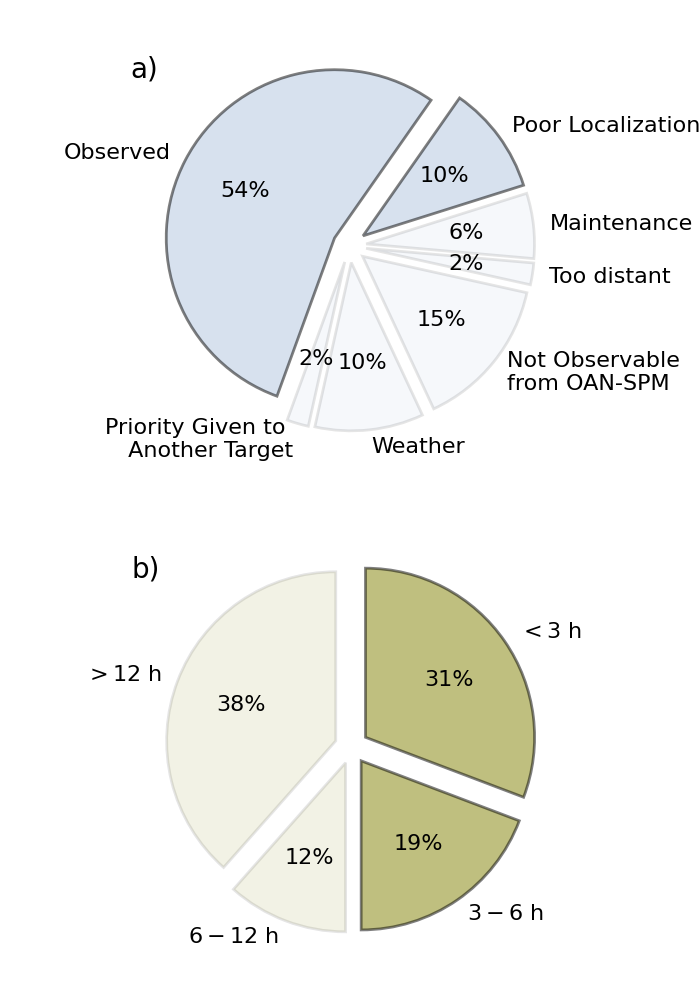}
 \caption{ (a) DDOTI response to unretracted alerts during the LIGO/Virgo O3 run. See Table~\ref{table:events} for details. (b) DDOTI reaction time to unretracted alerts,
 calculated as  the  time between the trigger and the first exposure (parameter observations $\Delta t$ of Table~\ref{tab:log}).}
 \label{fig:summarize}
\end{figure}

\section{DDOTI Observations}
\label{sec:ddoti}

\subsection{Hardware}

DDOTI (\url{http://ddoti.astroscu.unam.mx/}) is a wide-field, optical, robotic imager located at the Observatorio Astronómico Nacional (OAN) on the Sierra de San Pedro Mártir in Mexico \citep{Watson2016}.

DDOTI has an ASTELCO Systems NTM-500 mount with six Celestron RASA 28-cm astrographs in an ASTELCO Systems ARTS enclosure. Each telescope has an unfiltered Finger Lakes Instrumentation ML50100 front-illuminated CCD detector, an adapter of our own design and manufacture that allows static tip-tilt adjustment of the detector, and a modified Starlight Instruments motorized focuser. After repeated failures of the unmodified focusers, in September 2019 we installed focuser housings and direct couplings of our own design and manufacture, and these have proven to be reliable. 
Each telescope has a field of about $3.4\times3.4~\deg$ with 2.0 arcsec pixels. The individual fields are arranged on the sky in a $2 \times 3$ grid (nominally 6.8~deg E-W and 10.2~deg N-S) to give a total field of $69~\sqdeg$.

\subsection{Pipelines}
\label{sec:pipelines}

Our reduction pipeline subtracts dark images, performs iterative alignment taking into account atmospheric refraction and optical distortions, estimates and removes the background, resamples to a common pixel grid, performs clipping about the median to remove spurious data and satellite trails, and creates a variance-weighted coadded image. It uses {\sc sextractor} \citep{Bertin1996} for source detection, {\sc astrometry.net} \citep{Lang2010} for astrometry, and {\sc swarp} \citep{Bertin2010} for stacking and background estimation. 

Our photometry pipeline produces detections and magnitudes in the natural $w$ AB system of DDOTI.
It uses {\sc sextractor} \citep{Bertin1996} 
with a range of aperture diameters.
A weighted average of the flux in this set of apertures for all stars in a given subfield is then used to construct an annular point-spread function (PSF) whose core-to-halo ratio is allowed to vary smoothly over the field. This PSF is then fitted to the annular flux values for each source to optimize the signal-to-noise for point source photometry. The use of subfields accounts for the considerable variation of the PSF over the field of each CCD. The calibration is against the APASS DR10 catalog \citep{Henden2018} and uses our measured transformation of $w \approx r + 0.23 (g-r)$. The photometry pipeline determines a photometric normalization for each exposure and feeds this back into the reduction pipeline to iteratively correct transparency variations. Our final catalog contains photometry from both the individual frames and from the coadded image.

Our transients pipeline filters the detected source catalog to identify likely counterparts. It matches the sources against the USNO \citep{Monet2003}, PanSTARRS DR1 \citep{Tonry2012}, and SDSS DR9 \citep{Ahn2012} photometric catalogs. It eliminates sources within one FWHM of a USNO-B1 or APASS source, clustered detections (more than 2), those near very bright ($R<13$) catalog stars, those near any known minor planets whose positions are supplied by the Minor Planet Center's {\sc mpchecker} service, and those whose fluxes in partitions of the data are not consistent with their flux in the final image. We typically select sources at the $10\sigma$ level, since our experience is that our catalog filtering works well on null fields to this level but not to fainter levels. 

For events of particular interests, such as GW190814, image subtraction between different epochs was used to search for transient sources and this allows use to use a lower $6\sigma$ threshold \citep{Thakur2020}.

\begin{figure*}
\centering
 \includegraphics[width=0.8\textwidth]{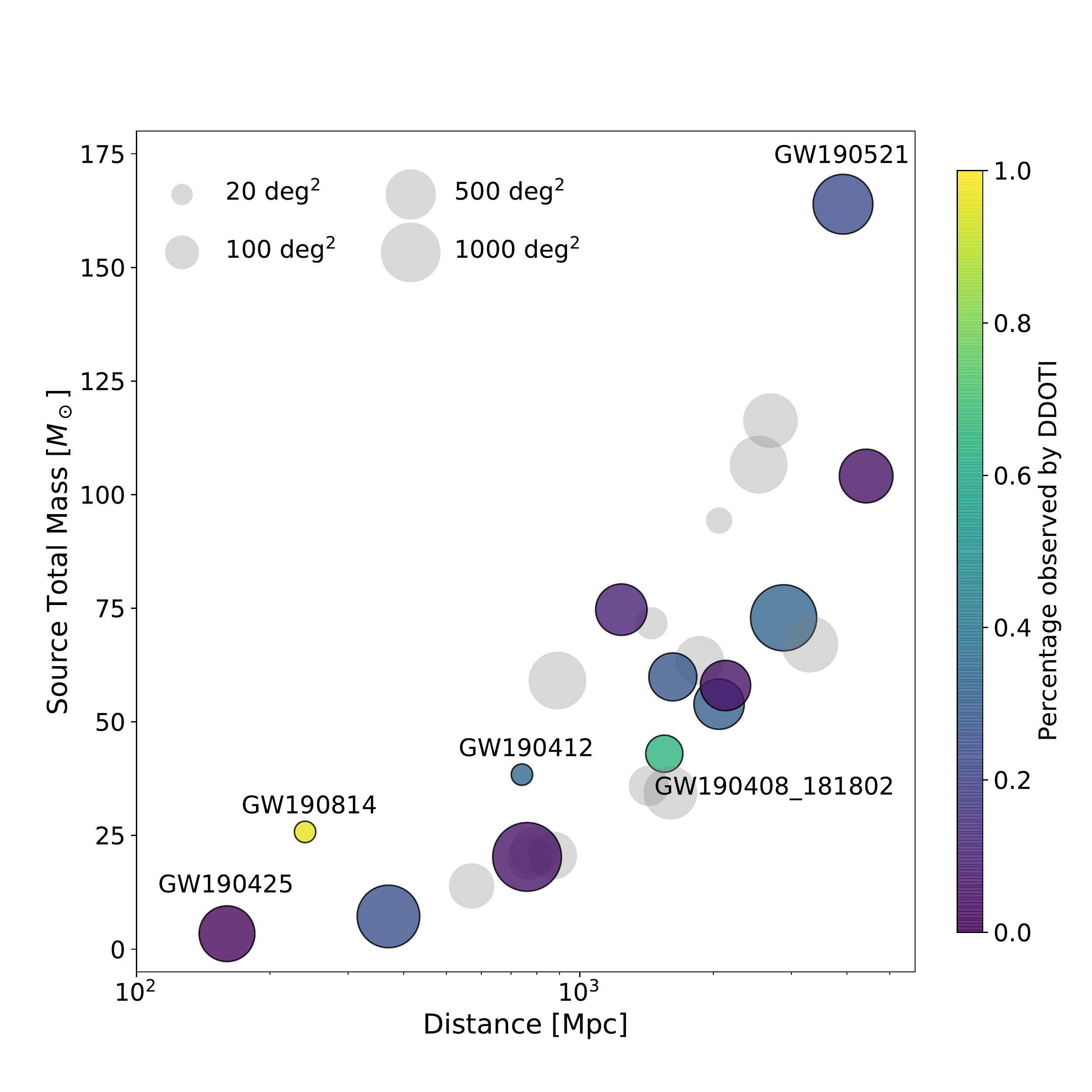}
 \caption{The distance and the source total mass for events during O3a. We use the information from the GWTC-2 catalog \citep{Abbott2020c}. The circle size is proportional to the sky localization and the colour shows the probability covered in the {\sc Bayestar/LALInference} maps by DDOTI. Grey circles illustrate the events that DDOTI did not observe. We label events that are of special interest.
 }
 \label{fig:massdistance}
\end{figure*}

\subsection{Observations}

\begin{table*}
	\centering
	\caption{LIGO/Virgo O3 Unretracted Real-Time Events.}
	\label{table:events}
  \begin{tabular}{llcl}
		\hline
				Event	&	Progenitor&DDOTI &Comments\\

		\hline
    GW190408$\_$181802& BBH  &$\checkmark$&\\
GW190412& BBH  &$\checkmark$&\\
GW190421$\_$213856& BBH   &$\checkmark$& Bad weather\\
GW190425& BNS  &$\checkmark$&\\
GW190426$\_$152155& NSBH  &$\checkmark$&\\
GW190503$\_$185404& BBH  &            &Too far south\\
GW190512$\_$180714& BBH  &            &Closed for maintenance\\
GW190513$\_$205428& BBH   &$\checkmark$&\\
GW190517$\_$055101& BBH   &            & Computer failed 20~m before event\\
GW190519$\_$153544& BBH  & &Bad weather (clouds)\\
GW190521& BBH   &$\checkmark$&\\
GW190521$\_$074359& BBH &$\checkmark$&\\
GW190602$\_$175927& BBH    &            &Too close to Sun\\
GW190630$\_$185205& BBH  &            &Closed for maintenance\\
GW190701$\_$203306& BBH   &            &Too close to Sun\\
GW190706$\_$222641& BBH   &$\checkmark$&\\
GW190707$\_$093326& BBH  &            &Bad weather (wind)\\
GW190720$\_$000836& BBH   &           & Poor initial localization\\
GW190727$\_$060333& BBH   &            &Bad weather (humidity)\\
GW190728$\_$064510& BBH   &            &Bad weather (humidity).\\
GW190814& MG &$\checkmark$&\\
GW190828$\_$063405& BBH  &$\checkmark$&\\
GW190828$\_$065509& BBH  &            & Priority given to GW190828$\_$063405\\
GW190915$\_$235702& BBH   &$\checkmark$&\\
GW190924$\_$021846& BBH   & & Close to Moon\\
GW190930$\_133541$& BBH    &$\checkmark$&\\
S191105e&BBH (95\%)&$\checkmark$&Alert delayed by about 28 h\\
S191109d&BBH (>99\%)&$\checkmark$&\\
S191129u&BBH (>99\%)&            &Poor initial localization\\
S191204r&BBH (>99\%)&$\checkmark$&\\
S191205ah&NSBH (93\%)&$\checkmark$&\\
S191213g&BNS (77\%)&            &Too close to Sun\\
S191215w&BBH (>99\%)&$\checkmark$&\\
S191216ap&BBH (99\%)&$\checkmark$&\\
S191222n&BBH (>99\%)&            &Bad weather (wind)\\
S200112r&BBH (>99\%)&            &Poor localization\\
S200114f& Burst        &$\checkmark$&\\
S200115j&MG (>99\%)&$\checkmark$&Bad weather (clouds)\\
S200128d&BBH (97\%)&            &Poor localization\\
S200129m&BBH (>99\%)&            &Too close to Sun\\
S200208q&BBH (99\%)&            &Too distant\\
S200213t&BNS (63\%)&$\checkmark$&\\
S200219ac&BBH (96\%)&$\checkmark$&\\
S200224ca&BBH (>99\%)&$\checkmark$&\\
S200225q&BBH (96\%)&$\checkmark$&\\
S200302c&BBH (89\%)&            &Poor localization\\
S200311bg&BBH (>99\%)&            &Too close to Sun\\
S200316bj&MG (>99\%)&$\checkmark$&\\
		\hline
	\end{tabular}
\end{table*}

During the O3 run, DDOTI observed 26 of the 48 events under consideration.  Table~\ref{table:events} lists these 48 events, indicates the 26 that DDOTI observed, and, for those that it did not, indicates why (see Figure~\ref{fig:summarize}).
DDOTI also observed the events S190910d and S190923y. However, we do not present these observations here since they were identified only in one detector in the GWTC-2 \citep{Abbott2020c} (see \S~\ref{sec:o3} for details).

We prioritized GW events over routine GRBs and other science observations. On nights in which we had more than one GW event, we prioritized events involving NS candidates, with good localizations, and with closer distances. DDOTI was closed from 2019 December 22 to 2020 January 6, due to the winter shutdown of the observatory, and from 2020 March 21, when the observatory closed in response to the COVID-19 health emergency.
 
Figure~\ref{fig:summarize} shows the observing efficiency and reaction time of DDOTI in response to O3 alerts.
Over half (26/48 or 54\%) of the real-time alerts were observed. 
Some other events which were potentially observable were not either because of their large localization errors (5/48 or 10\%) because priority was given to another GW source (1/48 or 2\%), or because they were considered to be too distant (1/48 or 2\%).
The remaining events were either not observable from the telescope's site (7/48 or 15\%), affected by poor weather conditions (5/48 or 10\%), or part of the small fraction (3/48 or 6\%) lost to maintenance or equipment failure. 

Figure~\ref{fig:summarize} also shows the delay between the event and the first DDOTI observations. The majority of DDOTI observations (61\%) started within 12 hours of the merger, and  approximately one third (32\%) within 3 hours. 

Even the DDOTI instantaneous field of 69~{\sqdeg} is small compared to a typical O3 uncertainty region, so mosaicing was necessary. We typically obtained about 1000 seconds of exposure on each field before moving to the next. In many cases, we were not able to cover the whole GW uncertainty region to this depth, and we typically preferred to reduce coverage rather than depth. 

Table~\ref{tab:log} shows the complete log of DDOTI observations of O3 events. For each event, we give the DDOTI coverage both in square degrees and the probability according to the {\sc Bayestar/LALInference} map, the delay between the event and receiving the alert (Alert $\Delta t$), the number of the night after the trigger, the delay between the event and the observations (Observations $\Delta t$), the total exposure time per field, the $10\sigma$ limiting magnitudes ($w_\mathrm{lim}$), the distance, the progenitor, and a reference our published GCN Circulars. 

DDOTI did not detect any likely EM counterpart to a GW event during the O3 run to our $10\sigma$ limiting magnitudes, which vary from event to event but are typically $w_\mathrm{lim}\approx18.5$--20.5 AB.

Figure~\ref{fig:massdistance} presents a summary of the DDOTI observations during O3a.
We use the GWTC-2 catalog \citep{Abbott2020c} to sort the sources according to their distance and total mass.  The symbol size is proportional to the sky localization area, and the color indicates the 
DDOTI probability coverage according to the {\sc Bayestar/LALInference} maps.

Figure~\ref{fig:observations} shows the DDOTI observations of events overlaid on the latest public {\sc Bayestar/LALInference} maps available from the {\sc GRACEDB} (\url{https://gracedb.ligo.org/superevents}) using the {\sc ligo.skymap} package (\url{https://pypi.org/project/ligo.skymap/}). For each event, orange squares represent the fields observed by DDOTI and the blue and purple shading shows the probability density in the {\sc Bayestar/LALInference} map. The positions of the Sun and Moon at the start of our observations are indicated by yellow and grey circles. The Figures also indicate the region of the sky available to DDOTI to a zenith distance of 72 degrees, which corresponds to an airmass of 3.2. The black dashed line encloses the region of the sky available to DDOTI at the start of the night (formally the start of evening astronomical twilight), the black dotted line encloses the region available to DDOTI at the end of the night (formally the end of morning astronomical twilight), and the black solid line encloses the region available at some point between the start and end of the night.

\begin{table*}
	\centering
	\caption{DDOTI Observations of LIGO/Virgo Events}
	\label{tab:log}
  \begin{adjustbox}{width=0.9\textwidth,center}
 \begin{tabular}{lrrrrrrccll} 
		\hline
			Event	&	\multicolumn{2}{c}{Coverage}&Alert $\Delta t$&	Night& Observations $\Delta t$	&Exposure	& $w_\mathrm{lim}$  & Distance&	Progenitor&GCN	\\
			&($\sqdeg$)&(\%)&(h)&&(h)&(h)&(10$\sigma$)&(Mpc)&\\
		\hline
    GW190408$\_$181802&121&64&0.6&1&16.8--17.8&0.7&18.3--18.9&1550&BBH&\cite{2019GCN.24086....1W}\\
GW190412&69&33&1&2&24.1--29.6&4&20.9--21.6&740&BBH\\
GW190421$\_$213856&400&32&18.8&2&30.8--37.4&3.6&18.7--19.2&2880&BBH\\
&&&&3&54.0--61.2&4.1&19.0--19.4&&\\
GW190425&80&2&0.7&1&1.6--2.9&0.6&19.1--20.4&160&BNS\\
&&&&2&23.2--27.5&2.2&20.6--21.2&&\\
GW190426$\_$152155&384&25&0.2&1&12.2--27.5&7.8&20.6--20.8&370&NSBH&\cite{2019GCN.24310....1W}\\
GW190513$\_$205428&80&30&0.5&1&12.3--13.1&0.4&19.3--19.5&2060&BBH\\
GW190521&280&23&0.1&1&1.5--7.0&3.4&20.0--20.2&3920&BBH&\cite{2019GCN.24644....1W}\\
GW190521$\_$074359&80&9&0.1&1&0.2--3.7&0.8&19.7--20.4&1240&BBH\\
GW190706$\_$222641&67&5&0.3&1&5.8--9.5&2.4&20.5--21.0&4420&BBH\\
GW190814&100&95&0.5&1&10.8--14.7&1.1&18.3--18.7&240&MG&\cite{2019GCN.25352....1D}\\
&&&&2&34.8--38.8&1.3&18.4--18.9&&\\
&&&&4&82.7--86.7&1.9&18.4--18.9&&\\
&&&&5&108.6--110.7&1.5&18.7--19.2&&\\
&&&&6&134.4--134.8&0.3&18.9--19.1&&\\
&&&&7&154.4--158.8&3&19.6--19.8&&\\
&&&&13&301.5--302.8&0.9&20.6--20.8&&\\
&&&&14&322.0--322.3&0.3&19.5--20.0&&\\
&&&&21&489.7--493.9&2.1&19.9--20.6&&\\
&&&&25&585.7--591.1&3.3&20.6--21.0&&\\
&&&&26&609.6--613.3&2.2&20.1--20.7&&\\
GW190828$\_$063405&231&5&0.3&1&1.4--5.5&2.5&20.0--20.2&2130&BBH&\cite{2019GCN.25562....1P}\\
&&&&2&20.7--22.4&1.2&20.0--20.1&&\\
&&&&4&68.8--77.4&4.4&19.1-20.2&&\\
&&&&5&92.8--99.7&3.6&19.8--20.0&&\\
GW190915$\_$235702&80&27&0.1&3&50.9--51.3&0.3&19.2--19.7&1620&BBH\\


GW190930$\_$133541&80&6&0.1&1&13.0--21.9&5.7&20.0--20.7&760&BBH\\
S191105e&80&6&28.4&9&207.2--207.5&0.3&18.0--19.0&1183&BBH\\
S191109d&310&15&0.2&1&6.5-9.1&3.2&19.8--20.0&1810&BBH\\
&&&&2&24.8-32.8&4&18.8--19.5&&\\
&&&&3&47.9-56.1&4&19.9--20.4&&\\
&&&&4&72.9-73.6&0.7&17.2--17.9&&\\
S191204r&465&94&0.7&1&15.0-15.3&0.3&18.3--19.1&678&BBH\\
S191205ah&132&31&0.1&1&3.8-6.2&2.4&18.2--18.8&385&NSBH\\
S191215w&59&4&0.1&1&3.2-5.1&1.9&18.8--20.4&1779&BBH\\
&&&&6&125.1-125.5&0.2&17.4--18.7&&\\
S191216ap&313&63&0.3&1&4.8-6.9&1.9&19.3--19.8&376&BBH\\
&&&&2&28.3--31.3&1.8&19.8–20.3&&\\
&&&&4&76.3--76.8&1.9&18.5–19.4&&\\
&&&&5&101.0--101.3&0.8&17.7–18.9&&\\
S200114f&280&80&0.1&1&0.5-9.7&1.1&19.6--20.1&999&Burst&\cite{2020GCN.26752....1D}\\
&&&&2&24.3-24.6&0.3&18.1--18.9&&\\
&&&&5&97.9-103.9&6&20.0--20.8&&\\
S200115j&116&25&0.1&1&1.3-1.6&0.3&17.2--18.4&340&MG\\
&&&&2&22.0-23.3&0.8&20.0--20.1&&\\
S200213t&200&37&0.1&1&0.6-2.6&1.6&18.7-19.8&201&BNS&\cite{2020GCN.27061....1W}\\
&&&&2&22.3-25.1&1.9&18.3--19.8&&\\
&&&&3&46.3-46.9&0.6&18.7--19.2&&\\
S200219ac&320&16&2.6&1&17.8-22.7&6.2&21.1--21.5&3533&BBH\\
S200224ca&135&92&0.1&1&7.1-9.8&4.7&20.7--21.3&1575&BBH&\cite{2020GCN.27212....1D}\\
S200225q&432&27&0.1&1&0.1-0.4&0.2&18.2--18.7&995&BBH\\
S200316bj&81&40&0.1&1&4.8-5.8&1.7&18.7--19.3&1778&MG&\cite{2020GCN.27402....1P}\\
		\hline

	\end{tabular}
 \end{adjustbox}
\end{table*}

\section{Analysis and Discussion}
\label{sec:results}

In this section we compare our upper limits on the optical emission from counterparts of the GW events with possible sources. We also look forward to the O4 run, for which the sensitivity of LIGO and Virgo is expected to be greater and the addition of the Kamioka Gravitational Wave detector (KAGRA) should significantly improve the precision and compactness of some of the localizations \citep{Abbott2020a}.

\subsection{Comparison to AT~2017gfo}
\label{sec:gfo}

The kilonova AT~2017gfo associated with the compact binary merger GW170817 remains the only spectroscopically confirmed kilonova to date.
Its optical emission was detected about 11~h after the merger
and was seen to fade rapidly \citep[e.g.,][]{Coulter2017,Arcavi2017,Evans2017,Troja2017}.
The near-infrared emission evolved on longer timescales, peaking a few days after the merger
\citep[e.g.][]{Pian2017,Troja2017,Tanvir2017}. 

\begin{figure*}
\centering
 \includegraphics[width=0.80\textwidth]{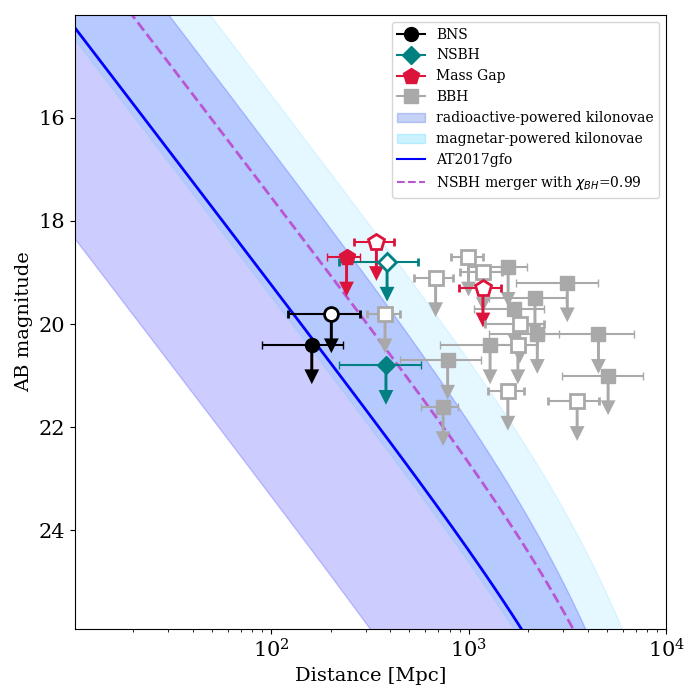}
 \caption{DDOTI deepest optical limits for the GW sources observed during the O3 run. 
The symbol shape and color indicates the merger type.
Events detected during O3a and confirmed in 
the GWTC-2 
are marked by the closed symbols.
Events detected during O3b are shown with open symbols. 
The blue solid line shows the peak magnitude of AT~2017gfo \citep{Arcavi2017} scaled by distance. 
The purple dashed line shows the kilonova model of \citet{Barbieri2020} for a NSBH merger with a maximally spinning BH. 
The shaded regions indicate the range of peak magnitudes spanned by radioactively powered (purple) and a magnetar-powered (blue) kilonovae, calculated using the models presented in \citet{Thakur2020} and \citet{2013Yu}, respectively.
}
 \label{fig:kilonova}
\end{figure*}

\begin{figure*}
\centering
 \includegraphics[width=0.80\textwidth]{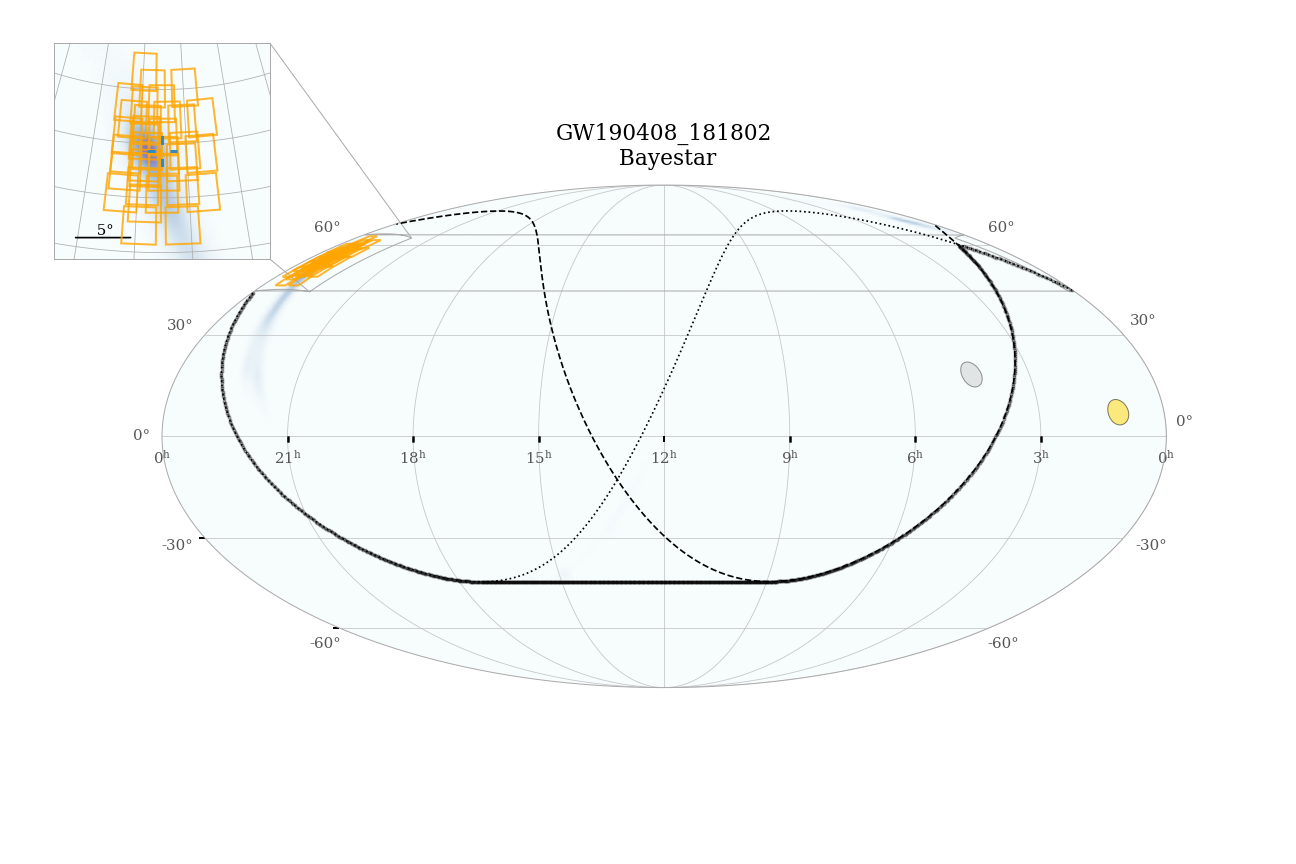}
 \caption{DDOTI observations of GW190408$\_$181802. DDOTI fields are shown as orange quadrilaterals and the LIGO/Virgo probability map by blue and purple shading. The Sun and Moon (at the moment when DDOTI began to observe) are indicated by yellow and grey circles respectively. The black dashed line indicates the region of the sky available to DDOTI at the start of the night, the black dotted line indicates the region available to DDOTI at the end of the night, and the black solid line indicates the region available at some point during the night.}
 \label{fig:observations}
\end{figure*}

In order to assess whether DDOTI observations are sensitive to a kilonova similar to AT~2017gfo, we compare our limits to the peak magnitude
$w = 17.49 \pm 0.04$ at 11~h reported by \cite{Arcavi2017}. We note that the $w$ band of \cite{Arcavi2017} is different from the DDOTI $w$ band, but to the accuracy we need here, the two can be treated as similar.
The comparison to AT~2017gfo is shown in Figure~\ref{fig:kilonova}, which reports 
the DDOTI first-night upper limits for each event observed during the O3 run. The arrows show the deepest $10\sigma$ limits obtained by the 6 cameras.
The predicted peak magnitude of AT~2017gfo as a function of the distance is shown by the solid line.

Figure~\ref{fig:kilonova} shows that only in the case of the nearby BNS GW190425 (at a distance of 160 Mpc) are our observations deep enough to have detected a kilonova like AT~2017gfo. However, the poor localization of 10,000~{\sqdeg} for this event \citep{Abbott2020c} means that our DDOTI observations include only 2\% of the probability and so we cannot constrain the properties of its EM emission.

Figure~\ref{fig:kilonova} shows that under favorable observing conditions DDOTI is able to detect kilonovae similar to AT~2017gfo out to about 200~Mpc. Clearly, obtaining good coverage of these rare events will be key, 
and for this better localizations from an expanded network of GW interferometers will be essential. 


\subsection{Comparison to Kilonova Models}
\label{sec:kilonova}

The brightness, color, and typical timescales of a kilonova depend on a variety of factors, such as the mass, spin, and stellar type of the progenitor stars, as well as the observer's viewing angle. 
A wide and diverse range of kilonova behaviours is to be expected, as also indicated by observations of sGRBs \citep[e.g.][]{Oconnor2021}.
For this reason, in Figure~\ref{fig:kilonova} we also explore DDOTI's ability to detect kilonovae different than AT~2017gfo. 

We consider two sets of models: kilonovae powered by the radioactive decay of freshly produced r-process nuclei and kilonovae powered by the spin-down energy of a newly formed magnetar. The $w$ magnitude was estimated by integrating the predicted spectrum over the SDSS-$r$ filter response and converting from $r$-band to the DDOTI $w$-band using the photometric transformation given in \S\ref{sec:pipelines}.

For kilonovae powered by radioactive decay, 
we estimate the kilonova peak magnitude from the light curves presented in \cite{Thakur2020} and derived from a large suite of numerical simulations performed at Los Alamos National Laboratory \citep{Wollaeger2018,Korobkin2021}. 
These simulations include the lanthanide opacities of \cite{Fontes2020} and span ejecta masses between 0.001 and 0.1 $M_{\odot}$, ejecta velocities between 0.05\,$c$ and 0.3\,$c$, and viewing angles between 0 (on-axis) and 90 (edge-on) degrees. Since DDOTI observes at optical wavelengths, we are only able to constrain the properties of  lanthanide-poor outflows with electron fraction  $Y_e>0.27$.
Figure~\ref{fig:kilonova} shows the range of peak magnitudes for these models as a function of distance with purple shading. 
Within the range of parameters explored by the LANL simulations, the diversity of kilonova behaviors leads to a spread in peak brightness of about 6 magnitudes, with some kilonovae being fainter than AT~2017gfo and others being brighter. 

On the other hand, a possible outcome of a BNS merger is a
rapidly spinning and highly magnetized NS, known as magnetar.  Recent measurements of NS radii and maximum mass \citep{Cromartie2020,Riley2019,Miller2019} favor a stiff EoS, predicting that a large fraction of BNS merger remnants
will be massive NSs. 
If the NS survives long enough, a significant amount of its rotational energy (up to $10^{53}$\,erg) can be transferred to the merger ejecta through its dipole spin-down radiation \citep[e.g.,][]{2013Yu}. 
The resulting kilonova would be bluer and brighter than its radioactively powered equivalent. 
This can be seen in Figure~\ref{fig:kilonova}, 
which shows in blue shading the predicted range of peak magnitudes for a kilonova with an initial NS magnetic field 
$B \sim 0.5 \times 10^{15}\,G$
and a NS lifetime of $10^{4}$ to $10^{5}$ s \citep[e.g.][]{2013Yu}. The absolute magnitudes of this magnetar-powered kilonovae span values from $M_r\sim$-15.6 to $M_r\sim$-19.5.

Considering Figure~\ref{fig:kilonova}, we can see that there are several events for which the depth of our observations and the distance to the event potentially allow us to constrain a kilonova powered by radioactive decay. Four of these events contained a low-mass compact object, possibly consistent with a NS. 
The first is the BNS merger GW190425, but as we have noted our coverage is only 2\%. The others are the candidate BNS merger S200213t, for which our coverage is 27\%, the candidate NSBH merger GW190426\_152155, for which our coverage is 25\%, and the MG event GW190814, already discussed in \citet{Thakur2020}. Our upper limits are compatible with a broad range of ejecta properties and viewing angles, although they disfavor mergers with large ejecta masses ($\gtrsim$0.1 $M_{\odot}$) seen closer to the polar axis.
Such massive ejecta are not expected from BNS mergers \citep{Fujibashi2020}, but, in favorable conditions, might be produced in NSBH encounters.

Turning to potential constraints on magnetar-powered kilonovae, the sample of relevant events is reduced to only the BNS mergers GW190425 and S200213t.
Although our limited coverage prevents us from drawing strong conclusions, Figure~\ref{fig:kilonova} demonstrates that DDOTI's sensitivity is sufficient to probe nearly the entire parameter space of bright magnetar-powered kilonovae. 

Figure~\ref{fig:kilonova} allows us to predict some possibilities for the O4 run. For the events out to about 100~Mpc,  DDOTI observations could probe a large portion of the parameter space, including kilonovae produced by low ejecta masses ($\approx0.03M_{\odot}$) or observed at large viewing angles ($>25$ deg). 
This is consistent with the findings of \citet{Chase2021}, who also estimate the peak detectability at 12 hours post-merger.
However, such nearby events are known to be rare \citep{Abbott2020c,Dichiara2020,Andreoni2021} and the majority of future GW sources will be found at much further distances. At distances between about 200 and 500 Mpc, DDOTI remains sensitive to luminous kilonovae powered by the radioactive decay of large ($\gtrsim$0.1\,M$_{\odot}$) ejecta masses and could place interesting constraints on the BH spin, in case of a candidate NSBH merger, and probe the nature of elusive MG objects. As an example, we also show in
Figure~\ref{fig:kilonova} the predicted peak brightness for a NSBH merger with chirp mass of approximately $1.4M_\odot$  and a maximally spinning ($\chi_{BH}=0.99$) solar-mass BH  \citep{Barbieri2020}.

Finally, in the case of an indefinitely stable NS powering the emission, the predicted signal would be so bright that it could be detectable up to 1~Gpc ($z \approx 0.2)$. Therefore, even in the cases of distant BNS mergers, DDOTI observations will be able to constrain the nature of the merger remnant.  
Concretely, a sample of 10 BNS mergers localized within about 100 {\sqdeg} would allow us to provide a robust test of the magnetar model and, if no bright kilonova is detected, place a tighter limit on the fraction of stable magnetars from BNS mergers.


\subsection{Comparison to sGRBs afterglows}
\label{sec:sgrb}

We can compare our limiting magnitudes to observed on-axis sGRBs  using the methodology of \cite{Watson2020}. Essentially, we assume that the observed optical flux density of an sGRB behaves as $F_\nu \propto t^{-\alpha}\nu^{-\beta}$, in which $t$ is the observed delay since the event and $\nu$ is the observed frequency. This leads to $F_\nu D^2t^\alpha(1+z)^{\beta-\alpha}$ being constant for a given sGRB, in which $D$ is the luminosity distance and $z$ is the redshift. We assume $\alpha= 1$ and $\beta=0.7$ \citep{Kumar2015}. These assumptions allow us to transform an observation of an sGRB at a given frequency, distance, and time to any other frequency, distance, and time.

We consider a sample of 36 sGRBs with known redshifts and either optical detections or upper limits. Our sample was compiled from  \cite{Watson2020}, \cite{Dichiara2021} and \cite{Oconnor2021}. 
We transform the magnitudes of the sample of sGRBs assuming the mean distance of each event, the median time of the earliest DDOTI observations, and the median frequency of the $w$ band. We can carry this out for all the O3 events except the burst S200114f, which has no distance measurement.  We then compare these values to the DDOTI upper limit. This gives us the raw probability that an sGRB from the sample would have been detected by DDOTI if it were associated with the event and if it occurred in the region covered by our observations. We treat sGRBs with upper limits in the same way as \cite{Watson2020}, calculating an optimistic raw probability under the assumption that the actual magnitude of the sGRB is just below the reported upper limit and a pessimistic raw probability under the assumption that the actual magnitude is too faint to be detected by DDOTI. We define the neutral raw probability to be the average of the optimistic and pessimistic raw probabilities. Finally, we multiply the neutral raw probability by the coverage to obtain the total detection probability.
Table~\ref{tab:calculations} shows for each event the average delay $t$ in hours, the average $w_{\rm lim}$ for our first night of observations, and the 
the total detection probability $\bar P_\mathrm{det}$.

For the BNS, NSBH, and MG events, the total detection probabilities range from 0.01 to 0.15, except for 0.47 for GW190814, which was already discussed by \cite{Watson2020} and \cite{Thakur2020}. 

A high total detection probability is favored for nearby events, observed quickly, with deep limiting magnitudes, and with good coverage. Unfortunately, these circumstances did not occur with regularity in O3 and only 4 of the 25 events have total detection probabilities of more than 15\%. There are several events which have raw detection probabilities of order 50\% (GW190412, GW190425, GW190426\_152155, GW190814, GW190930\_133551, S191216ap, and S200213t), but in many cases the relatively low coverage reduces the total detection probability significantly.



\begin{table*}
	\centering
	\caption{Probabilities of Detection and Energies Calculated for Expected sGRBs }
	\label{tab:calculations}
 \begin{tabular}{lrrrr} 
		\hline
			Event & $t$&$w_\mathrm{lim}$& $\bar{P}_\mathrm{det}$&$E_{\rm K,iso,52}$\\
		\hline
    GW190408$\_$181802&17.30&18.60&0.04&186.39 -- 285.13\\
GW190412&26.85&21.25&0.14&42.28 -- 69.42\\
GW190421$\_$213856&34.10&18.95&0.01&205.61 -- 293.01\\
GW190425&2.25&19.75&0.01&19.86 -- 49.88\\
GW190426$\_$152155&19.85&20.70&0.14&64.71 -- 74.56\\
GW190513$\_$205418&12.70&19.40&0.03&92.05 -- 106.06\\
GW190521&4.25&20.10&0.01&21.67 -- 24.97\\
GW190521$\_$074359&1.95&20.03&0.01&14.86 -- 24.39\\
GW190706$\_$222641&7.65&20.75&$<0.01$&17.71 -- 25.24\\
GW190814&12.75&18.50&0.47&216.29 -- 287.16\\
GW190828$\_$063405&3.45&20.10&$<0.01$&22.55 -- 25.98\\
GW190915$\_$235702&51.10&19.45&0.03&221.80 -- 316.09\\
GW190930$\_$133541&17.45&20.02&0.05&59.15 -- 97.13\\
S191105e&207.35&18.50&0.01&2076.17 -- 1022.28\\
S191109d&7.80&19.90&0.02&47.59 -- 54.83\\
S191204r&15.15&18.70&0.10&168.98 -- 297.84\\
S191205ah&5.00&18.50&0.09&102.38 -- 156.62\\
S191215w&4.15&19.60&$<0.01$&23.28 -- 72.32\\
S191216ap&5.85&19.55&0.27&56.30 -- 80.24\\
S200115j&1.45&18.20&0.05&58.23 -- 77.31\\
S200213t&1.60&19.25&0.15&23.78 -- 51.85\\
S200219ac&20.25&21.30&0.01&26.35 -- 34.98\\
S200224ca&8.45&21.00&0.23&20.66 -- 31.61\\
S200225q&0.25&18.45&0.01&12.42 -- 17.69\\
S200316bj&5.30&19.00&0.04&65.34 -- 99.95\\
		\hline
	\end{tabular}
\end{table*}

We note that the sample of observed sGRBs may well be biased in favour of brighter afterglows.
We used a set of simulations to explore a plausible range of afterglow parameters and independently assess whether DDOTI could detect on-axis afterglow emission or not. We used \textsc{afterglowpy} \citep{Ryan2020} to simulate 10,000 light-curves assuming a Gaussian-shaped jet with an opening angle $\theta_c = 4$ deg, as measured in GW170817 \citep{Troja19}, and log-normal distributions for the isotropic equivalent kinetic energy (log($E_{0}$), with a mean of 51.5 and $\sigma=1$), the density of the external medium (log($n$), with a mean of $-2.0$ and $\sigma=1$; \citealt{Oconnor2020}), and the fraction of energy transferred to the electrons (log($\epsilon_{e}$), with a mean of $-1.0$ and $\sigma=0.3$; \citealt{BeniaminivanderHorst}). 
The fraction of energy transferred to the magnetic field ($\epsilon_{B}$) 
is only loosely constrained by observations \citep{Santana14}, and we assumed a uniform distribution of log(($\epsilon_{B}$) between $-4$ and $-1$.
For the electron energy distribution $N(E) \propto E^{-p}$ we fixed the spectral index at $p=2.2$.

Assuming a horizon distance of 200~Mpc, as expected for BNS mergers, we found that about 50\% of the on-axis afterglows could be detected by DDOTI if observed within 12 hr in good conditions ($w_{\rm lim}=20.5$ mag) and with complete coverage.  These numbers increase to over 70\%  for events observed within 3 hrs after the merger. If we instead consider poor conditions ($w_{\rm lim}=17.9$ mag), the percentage of detectable on-axis afterglows decreases to less than $40$\%. 

Mergers of more massive objects (e.g., BBH and NSBH systems) are found by GW detectors at farther distances, and their afterglows are more challenging to detect even when the GRB jet is directed along our line of sight. Assuming a horizon distance of 1 Gpc, the percentage of optical afterglows visible by DDOTI would decrease by about half.
Our estimates conservatively consider standard forward shock emission from the GRB blastwave and would be higher in case of bright reverse shock emission, as recently detected in a growing number of sGRBs \citep[e.g.,][]{Becerra2019,Becerra2021}. 

We also tested our ability to detect possible off-axis events by simulating a distribution of viewing angles similar to the one expected for detected gravitational-wave signals \citep{Schutz2011}. In this case, the fraction of detectable sGRB afterglows drops below 5\%. Nevertheless, it is important to mention that LVC is most sensitive to mergers viewed off-axis \citep[see Figure 4][]{Schutz2011}.

Once again, looking forward to O4, we see that DDOTI has the sensitivity to detect sGRB afterglows out to the expected BNS horizon, but the key will be to increase the coverage of the GW localization and thereby the total detection probability. 

\subsection{Constraints on the Associated Blast Waves}
\label{sec:bbh}

In this section, we estimate the constraints on the parameters of a blast wave produced by the expected sGRBs described in \S~\ref{sec:sgrb}. We use equation 20 of \cite{Wang2015} (valid in general for a spherical blast wave), assuming  $\epsilon_E=0.1$, $\epsilon_B=0.01$, and $p=2.2$ as the microphysical parameters of the synchrotron emission process, and a density $n=10^{-3}$ cm$^{-3}$.
We convert the upper limits on the magnitude of the putative counterparts into limits on the isotropic kinetic energy  $E_\mathrm{K,iso,52}$ of any associated spherical blast-wave. We assume $\nu=10^{14}$~Hz, take $t_d$ to be the median time of the observations with DDOTI on the first night, and use the upper limits on the flux and the distances of the GWs as reported in Table~\ref{tab:log}.
The values of $E_\mathrm{K,iso,52}$ are given in the last column of Table~\ref{tab:calculations}.

For the closest events, the upper limits correspond to a maximum isotropic energy of $E_{\rm K,iso} \lesssim 2-3 \times 10^{53}$ ergs. 
These upper limits on the kinetic energy exclude the high-luminosity tail of sGRBs seen on-axis (which have isotropic energies of order  $10^{51}-10^{54}$~erg).  We cannot exclude the presence of lower-luminosity on-axis sGRBs or  off-axis GRBs \citep{Urrutia2021}, where the luminosity drops quickly with the observing angle (see, e.g., Figure 2 of \citealt{Granot2018}), and which are expected to dominate the GRB population. This is consistent with the 4/25 events with an on-axis detection probability above 15\% estimated in \S~\ref{sec:sgrb}.

\subsection{AGN Flares and BBHs}
\label{sec:gw190521}

As Figure~\ref{fig:massdistance} shows, the population of detected GW sources is dominated by BBH mergers at a median distance of about 2 Gpc. The EM signatures of these systems are uncertain, and predictions span a wide range of luminosities, spectral properties, and timescales \citep[e.g.][]{Perna2016,Zhang2016,Yi2019,McKernan2020}.
Very interestingly, GW observations rule out a simple power-law mass spectrum for these systems, and favor multi-component models, such as a power-law with one or more peaks \citep{LIGOpop}. An intriguing possibility consistent with these findings is that some of these high-mass BBH systems form through alternative channels, such as hierarchical mergers within an AGN accretion disk \citep{Jiang2019}. A gaseous environment such as this could provide the necessary fuel to power either a prompt EM counterpart \citep{Bartos2017,Yi2019} or a delayed off-nuclear flare emerging weeks to months after the merger
\citep{McKernan2020}.

\citet{Graham2020} proposed a plausible EM candidate counterpart of GW190521, the most massive BBH merger produced by the merger of an $85 M_{\odot}$ BH and a $66 M_{\odot}$ BH. They reported 
a rapid rise in the light curve of the AGN J124942.3+344929 \citep[see Figure 2 of ][]{Graham2020} detected by the Zwicky Transient Facility (ZTF) 34 days after the BBH merger. The photometry shows a rise from $r \approx 18.9$ to $r \approx 18.6$ followed by a fall over more than 100 days.

DDOTI observed GW190521 and covered the LIGO probability peak in the northern hemisphere, including the position of J124942.3+344929. The source is detected in our image at a magnitude of $w=18.9$, which is consistent with the catalogued value and therefore the source was not flagged as a potential counterpart by our transients pipeline. 
However, our search was limited to one night and could only recover prompt flaring events, rather than the month-long variability discussed in \citet{Graham2020}. 

The case of GW190521, in the end, was not optimal for EM searches with DDOTI due to its unfavorable localization, largely in the southern hemisphere, and its large distance. The probability of a chance alignment between AGN J124942.3+344929 and GW190425 is also not negligible \citep{Estelle2021}.

Predictions for the fourth observing run (O4) appear more promising for testing the association between BBH mergers and AGN flares. 
Based on a BBH merger rate of approximately $25\,\mathrm{Gpc^{-3} yr^{-1}}$ and the BH mass spectrum \textit{power-law + peak} \citep{LIGOpop}, 
we expect a large population of merging massive BBH systems, with primary components more massive than $50M_\odot$, within the horizon distance $z<0.3$. If we assume a mass-ratio distribution proportional to $q^{1.4}$, a significant fraction (about 40\%) of these mergers would have a total mass heavier than $100 M_{\odot}$ and represent an excellent test of alternative BH formation channels. 

In order to assess how many of these systems could be detected by the advanced LIGO/Virgo network and observed by DDOTI, we ran a basic injection test using the LSC Algorithm Library Suite (LALSuite) routines with a \textsc{IMRPhenomTPHM} waveform, then retrieved the detected coincidences using BAYESTAR \citep{Singer2016} and the predicted noise sensitivity curves for O4 (aLIGOAdVO4IntermediateT1800545 and aLIGOAdVO4T1800545).
Our simulation is rather conservative as it does not include the contribution of the Kamioka Gravitational Wave detector (KAGRA), also expected to be operative during O4. We assume a 70\% duty cycle for each detector and, since we are interested in well-localized sources, only consider events triggered by a minimum of two detectors and with a combined signal-to-noise ratio of at least 12.

We predict that 10--20 coincident detections will meet these criteria. Of these, about 60\% are predicted to be localised within 150 {\sqdeg} (90\% credible region) and within a median volume of 0.05 Gpc$^{-3}$, a factor about 200 smaller than the volume enclosed by GW190521. Such a localisation, although still large, would allow for a more confident investigation of the possible association between BBH mergers and AGN activity. 
Assuming the same sky distribution of events detected during O3, DDOTI would be able to observe and provide long-term monitoring for about 60\% of these targets (3--6 $\mathrm{yr}^{-1}$). 
However, the public release of preliminary mass estimates will be essential to classify these events as high-mass BBH mergers and to identify them as priority targets for long-term monitoring. 

Moreover, \citet{Zhang2019} discussed the possibility of observing radiation from gravitationally accelerated BHs with electric charge (dominated by the electric dipole).
We note that this process is similar to that of charges radiating by synchrotron emission (replacing electrically charged particles with charged BHs and the magnetic field by a gravitational field or space-time curvature). Thus, the radiation is concentrated towards the orbital plane (i.e., in directions perpendicular to the direction of propagation of any relativistic jet). Nonetheless, given that the BHs reach only slightly relativistic velocities of order $c/2$, the relativistic beaming (along the orbital plane) would be small.  This radiation would be mainly visible during the late stages of the inspiral phase and, perhaps, during the ring-down.  These two considerations guide us to assume that such emission would not have been detectable by DDOTI, as it observed minutes to hours after the merger.

\section{Summary}
\label{sec:summary}

We used DDOTI to obtain optical observations of 26 real-time GW events in the LIGO/Virgo O3 run, typically covering 20--40\% of the probability in {\sc Bayestar/LALInference} maps within less than 12 hours of the alerts. We detected no likely EM counterparts, but nevertheless obtained useful upper limits.

We have compared our upper limits with the several possible counterparts: a kilonova like AT~2017gfo; a wide range of models for kilonovae powered by radioactive decay or by a magnetar; an observed sample of on-axis sGRB afterglows; and models for on-axis and off-axis sGRB afterglows. While in some cases we have sufficient sensitivity to potentially restrict these scenarios, the large positional uncertainties during O3 lead to insufficient coverage in almost all cases and this prevents us from drawing conclusions at this moment.

Looking forward to O4, the improved sensitivity of existing detectors and the addition of the KAGRA detector should significantly improve the localization of some events, both in terms of the total area and the distribution \citep{Abbott2020a}. 
This will allow three improvements over our observations during the O3 campaign. First, with better localizations we expect to have more complete coverage and so higher probabilities of actually having observed the position of the GW source. Second, with more compact localizations we expect to be able to implement a strategy that cycles between fields with a higher cadence. For example, DDOTI can observe a $20 \times 20$~deg region with a cadence of about 8 minutes and a $14 \times 10$~deg region with a cadence of about 2.5 minutes. This makes it easier to detect rapidly fading or varying sources. Finally, with better localizations we expect to be able to obtain deeper exposures, since we will need to cover less sky. All of these will improve our ability to draw conclusions from our observations. For example, in 2.5 hours, DDOTI routinely observes 400~{\sqdeg} (i.e., six fields covering a region of $20 \times 20$~deg) with an exposure of 1000~s and obtains a $10\sigma$ limiting magnitude of $w = 18.5$ to 20.5, depending on conditions. If the uncertainty is reduced to about 100~{\sqdeg} (i.e., two fields), in the same time DDOTI can obtain three times as much exposure and improve the limiting magnitudes by about 0.6 mag. 

 We are also looking to improve our transients pipeline to routinely compare to prior reference images. Relative to our existing approach of comparing our detections against catalogs, this will improve the confidence with which we can identify new sources and allow us to reduce our limiting magnitude from $10\sigma$ to perhaps $6\sigma$, and thereby also gain about 0.5 mag \citep[as we did in][]{Thakur2020}. It will also allow us to reject previously detected transients (e.g., supernovae) and significantly improve our ability to detect transient sources close to galaxies.

We have two specific predictions for O4. First, if we can obtain good limits and coverage on about 10 BNS mergers within the detector horizon of about 200~Mpc, we should be able to either detect the emission of magnetar-powered kilonovae or eliminate these as viable models. Second, DDOTI will be able to provide long-term monitoring of the localization regions of high-mass BBH mergers to investigate a possible associate with AGN flares. However, this would require the prompt release of information on the nature of the BBH mergers, possibly the estimated chirp mass or simply a flag to indicate that the merger probably involves at least one object with a mass of at least 50 M$_\odot$.

\begin{figure*}
\centering
 \includegraphics[width=0.80\textwidth]{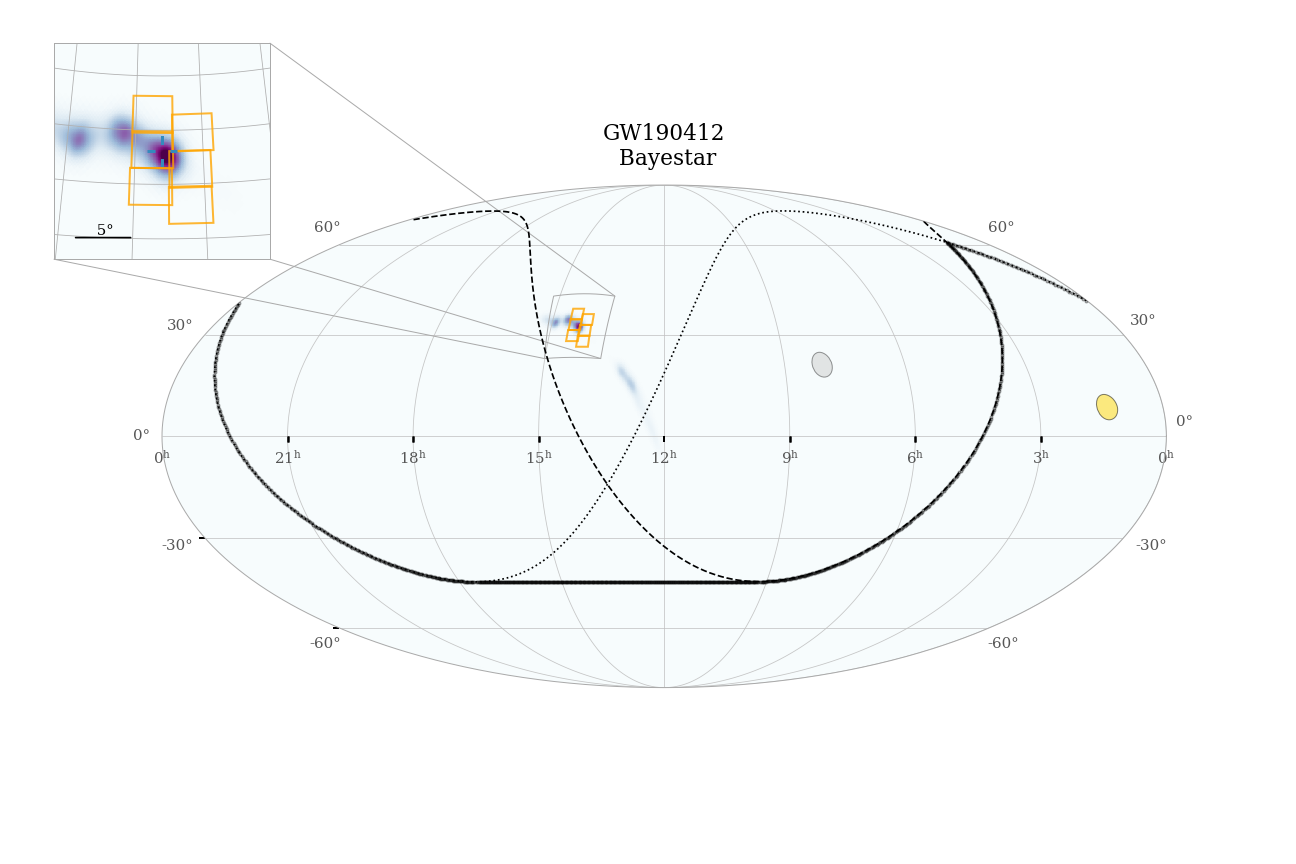}
 \caption{DDOTI observations of GW190412. DDOTI fields are shown as orange quadrilaterals and the LIGO/Virgo probability map by blue and purple shading. The Sun and Moon (at the moment when DDOTI began to observe) are indicated by yellow and grey circles respectively. The black dashed line indicates the region of the sky available to DDOTI at the start of the night, the black dotted line indicates the region available to DDOTI at the end of the night, and the black solid line indicates the region available at some point during the night.}
 \label{fig:observations1}
\end{figure*}

\begin{figure*}
\centering
 \includegraphics[width=0.80\textwidth]{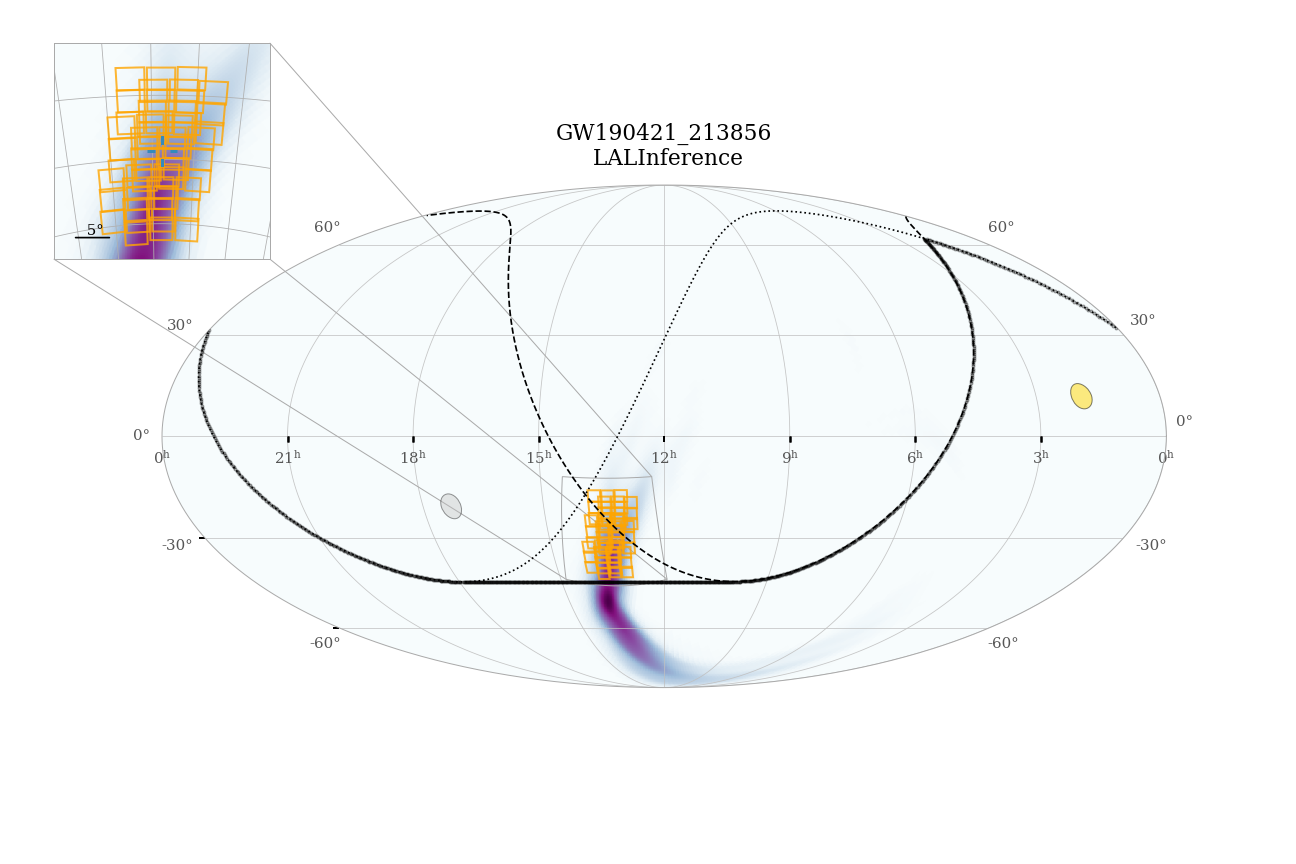}
 \contcaption{DDOTI observations of GW190421$\_$213856. DDOTI fields are shown as orange quadrilaterals and the LIGO/Virgo probability map by blue and purple shading. The Sun and Moon (at the moment when DDOTI began to observe) are indicated by yellow and grey circles respectively. The black dashed line indicates the region of the sky available to DDOTI at the start of the night, the black dotted line indicates the region available to DDOTI at the end of the night, and the black solid line indicates the region available at some point during the night.}
 \label{fig:observations2}
\end{figure*}

\begin{figure*}
\centering
 \includegraphics[width=0.80\textwidth]{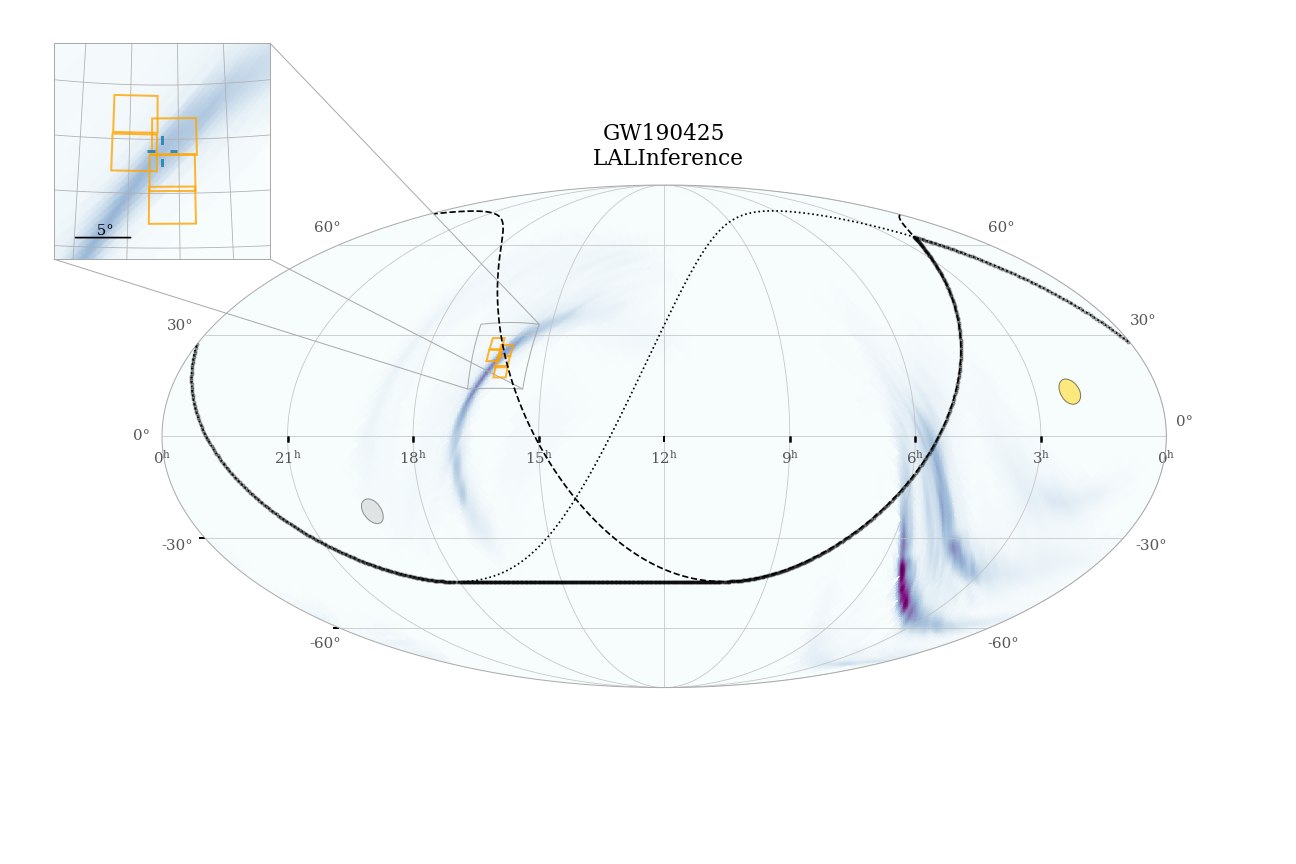}
 \contcaption{DDOTI observations of GW190425. DDOTI fields are shown as orange quadrilaterals and the LIGO/Virgo probability map by blue and purple shading. The Sun and Moon (at the moment when DDOTI began to observe) are indicated by yellow and grey circles respectively. The black dashed line indicates the region of the sky available to DDOTI at the start of the night, the black dotted line indicates the region available to DDOTI at the end of the night, and the black solid line indicates the region available at some point during the night.}
 \label{fig:observations3}
\end{figure*}

\begin{figure*}
\centering
 \includegraphics[width=0.80\textwidth]{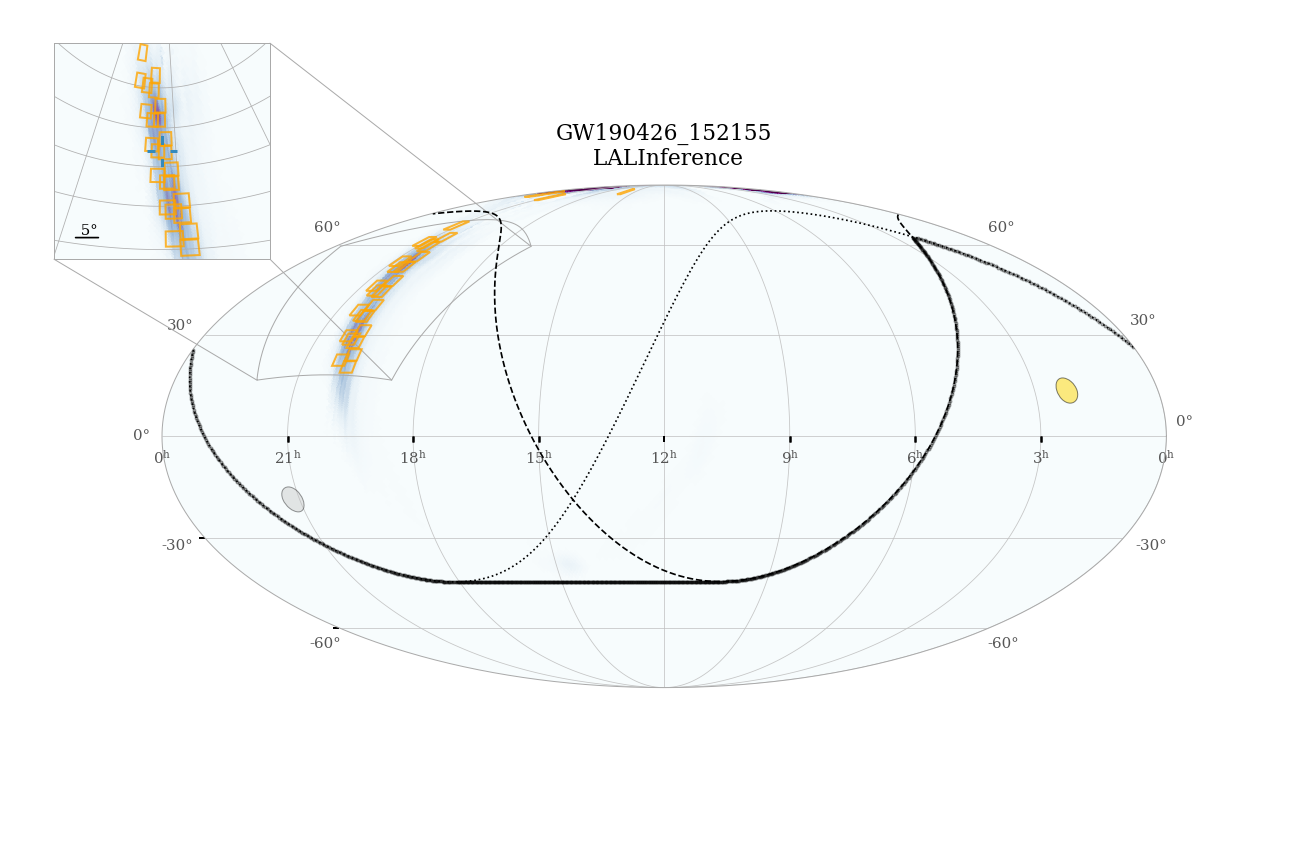}
 \contcaption{DDOTI observations of GW190426$\_$152155. DDOTI fields are shown as orange quadrilaterals and the LIGO/Virgo probability map by blue and purple shading. The Sun and Moon (at the moment when DDOTI began to observe) are indicated by yellow and grey circles respectively. The black dashed line indicates the region of the sky available to DDOTI at the start of the night, the black dotted line indicates the region available to DDOTI at the end of the night, and the black solid line indicates the region available at some point during the night.}
 \label{fig:observations4}
\end{figure*}

\begin{figure*}
\centering
 \includegraphics[width=0.80\textwidth]{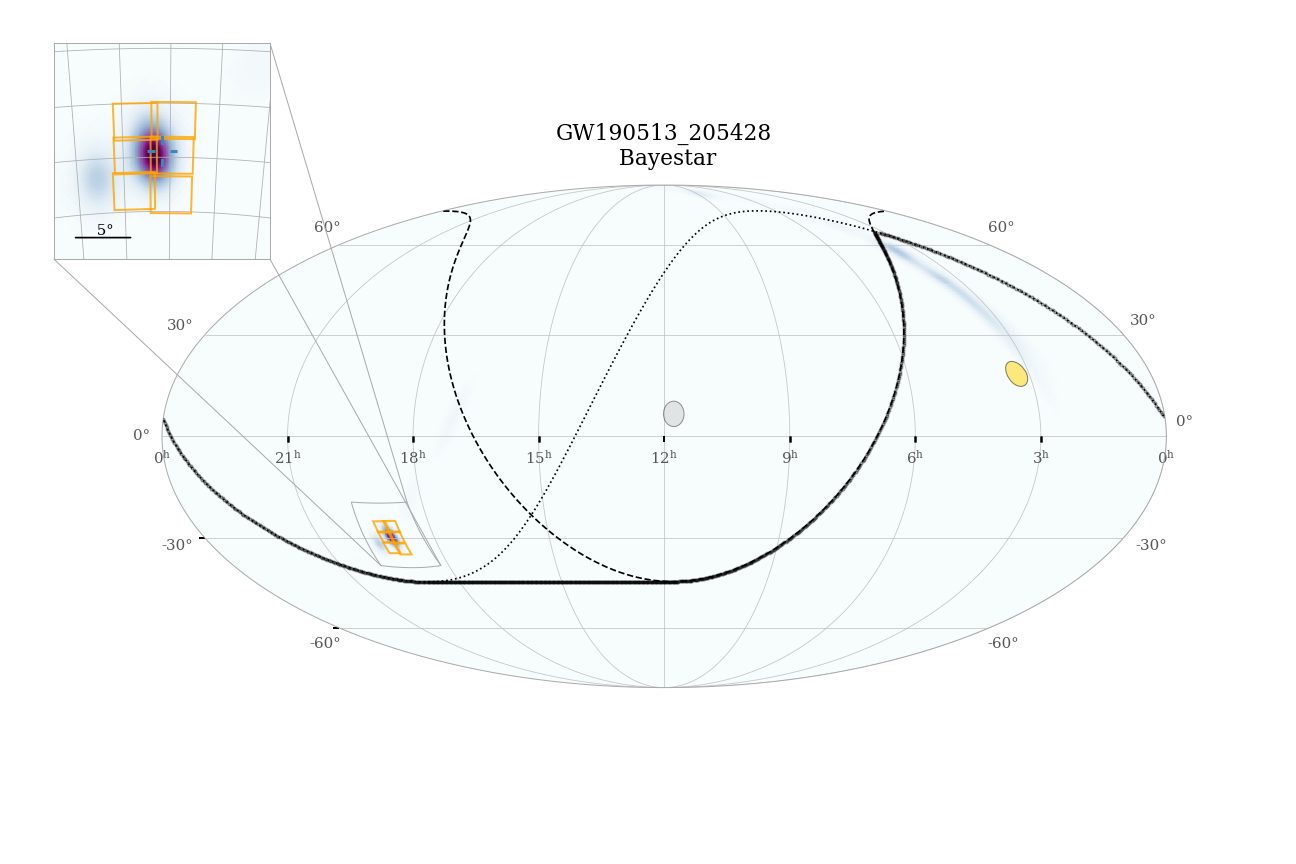}
 \contcaption{DDOTI observations of GW190513$\_$205428. DDOTI fields are shown as orange quadrilaterals and the LIGO/Virgo probability map by blue and purple shading. The Sun and Moon (at the moment when DDOTI began to observe) are indicated by yellow and grey circles respectively. The black dashed line indicates the region of the sky available to DDOTI at the start of the night, the black dotted line indicates the region available to DDOTI at the end of the night, and the black solid line indicates the region available at some point during the night.}
 \label{fig:observations5}
\end{figure*}

\begin{figure*}
\centering
 \includegraphics[width=0.80\textwidth]{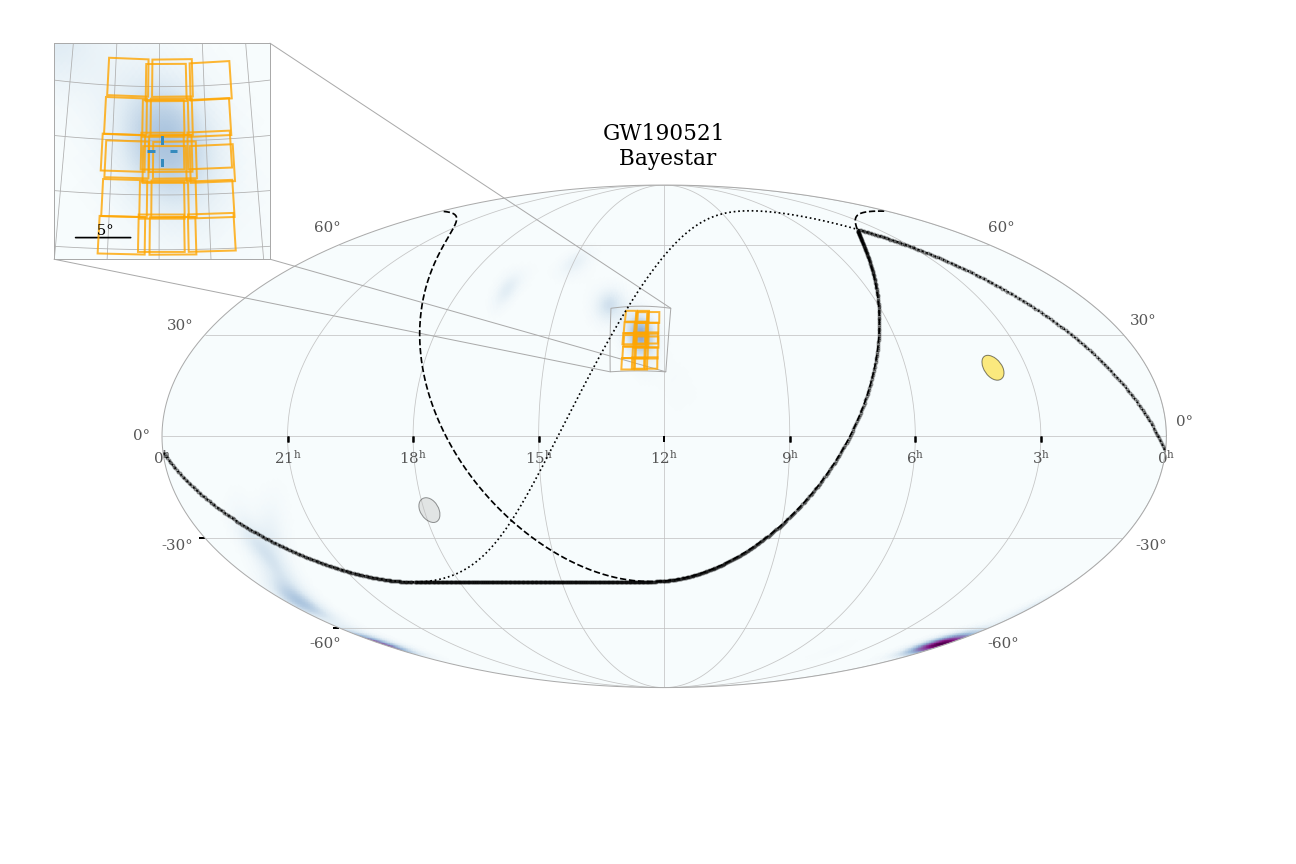}
 \contcaption{DDOTI observations of GW190521. DDOTI fields are shown as orange quadrilaterals and the LIGO/Virgo probability map by blue and purple shading. The Sun and Moon (at the moment when DDOTI began to observe) are indicated by yellow and grey circles respectively. The black dashed line indicates the region of the sky available to DDOTI at the start of the night, the black dotted line indicates the region available to DDOTI at the end of the night, and the black solid line indicates the region available at some point during the night.}

 \label{fig:observations6}
\end{figure*}

\begin{figure*}
\centering
 \includegraphics[width=0.80\textwidth]{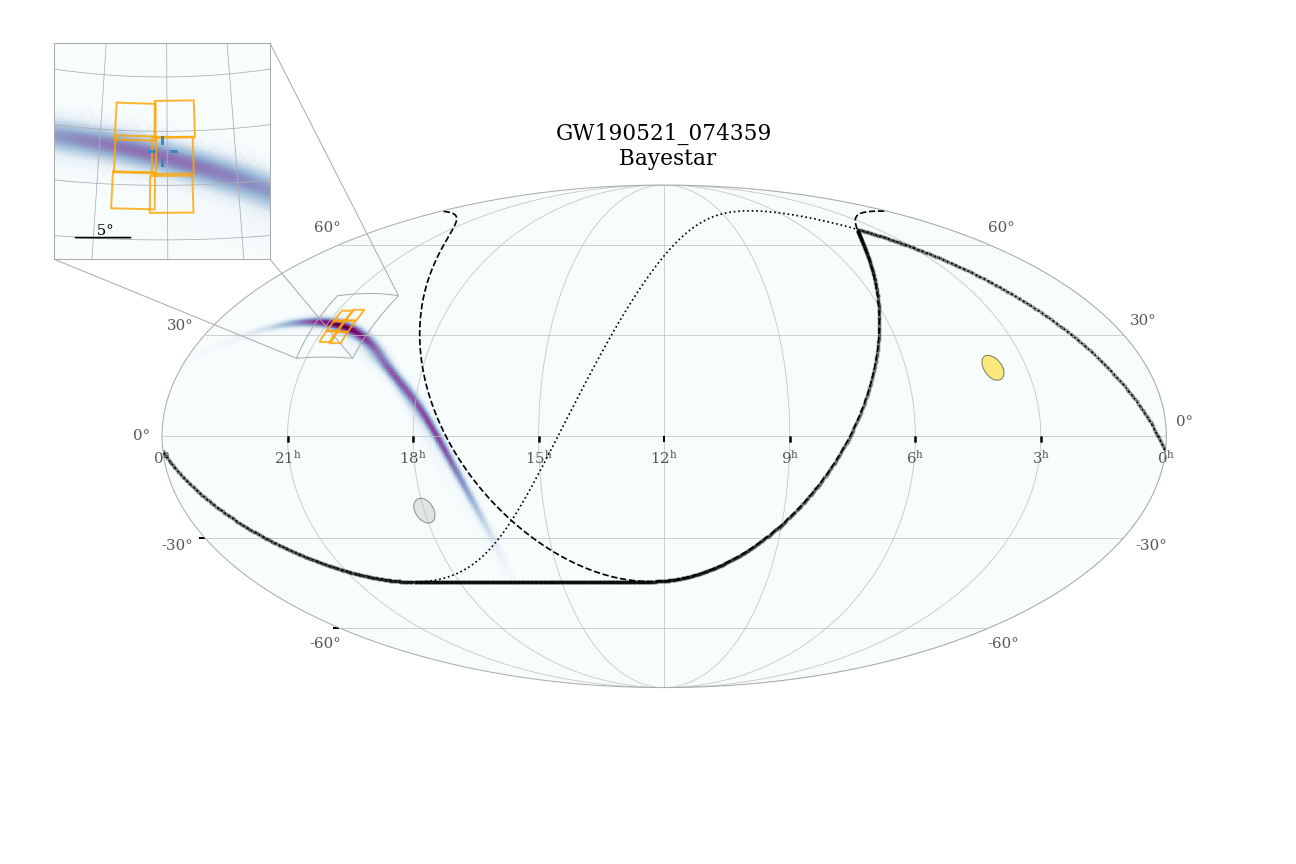}
 \contcaption{DDOTI observations of GW190521$\_$074359. DDOTI fields are shown as orange quadrilaterals and the LIGO/Virgo probability map by blue and purple shading. The Sun and Moon (at the moment when DDOTI began to observe) are indicated by yellow and grey circles respectively. The black dashed line indicates the region of the sky available to DDOTI at the start of the night, the black dotted line indicates the region available to DDOTI at the end of the night, and the black solid line indicates the region available at some point during the night.}
 \label{fig:observations7}
\end{figure*}

\begin{figure*}
\centering
 \includegraphics[width=0.80\textwidth]{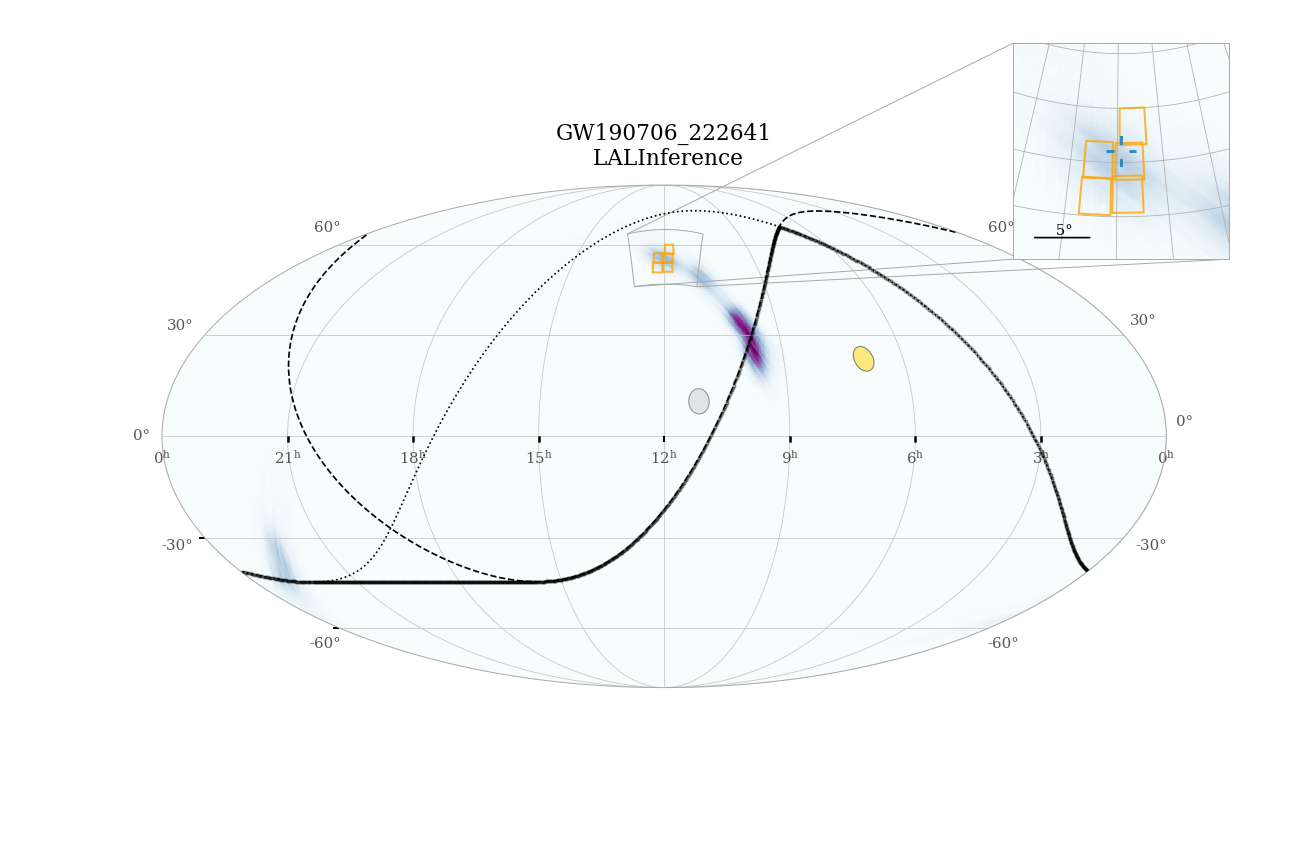}
 \contcaption{DDOTI observations of GW190706$\_$222641. DDOTI fields are shown as orange quadrilaterals and the LIGO/Virgo probability map by blue and purple shading. The Sun and Moon (at the moment when DDOTI began to observe) are indicated by yellow and grey circles respectively. The black dashed line indicates the region of the sky available to DDOTI at the start of the night, the black dotted line indicates the region available to DDOTI at the end of the night, and the black solid line indicates the region available at some point during the night.}

 \label{fig:observations9}
\end{figure*}

\begin{figure*}
\centering
 \includegraphics[width=0.80\textwidth]{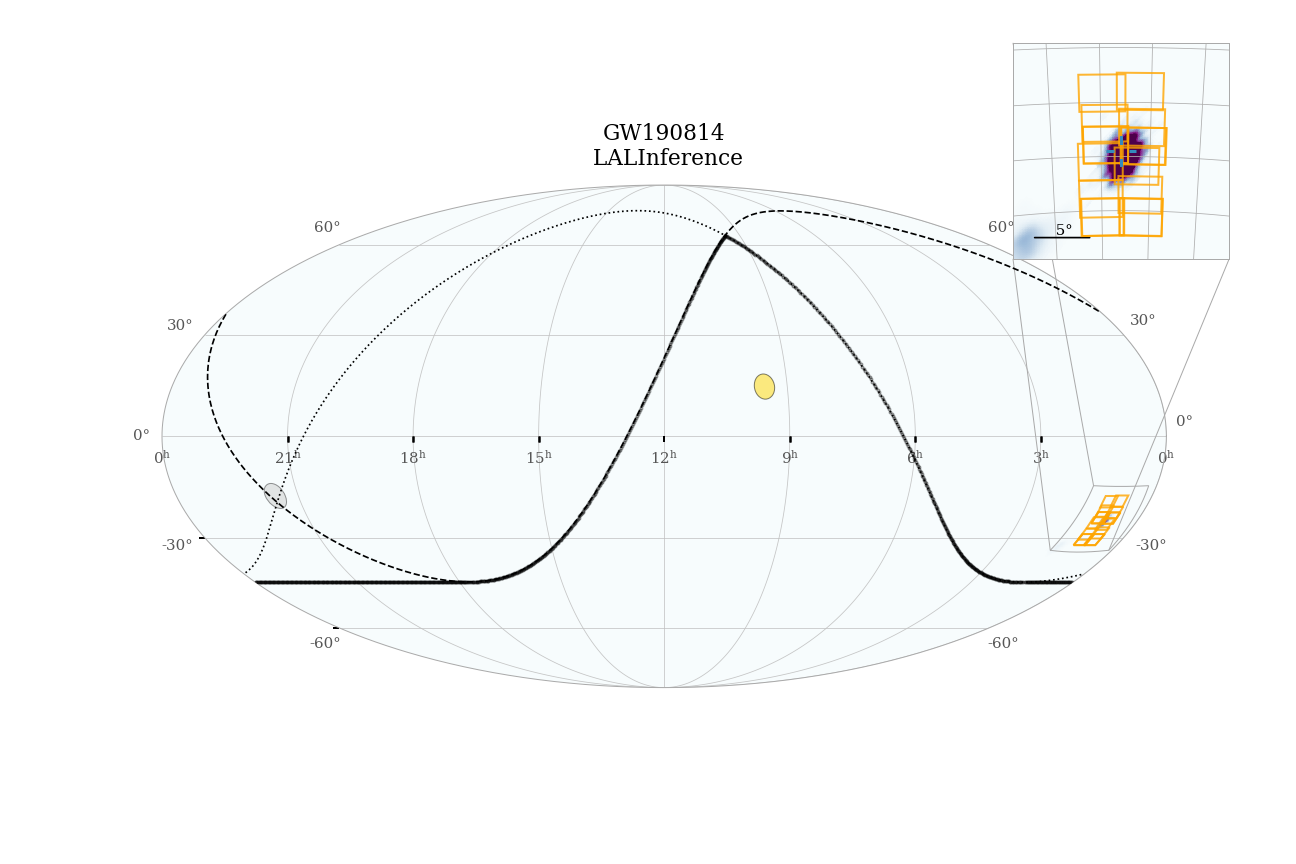}
 \contcaption{DDOTI observations of GW190814. DDOTI fields are shown as orange quadrilaterals and the LIGO/Virgo probability map by blue and purple shading. The Sun and Moon (at the moment when DDOTI began to observe) are indicated by yellow and grey circles respectively. The black dashed line indicates the region of the sky available to DDOTI at the start of the night, the black dotted line indicates the region available to DDOTI at the end of the night, and the black solid line indicates the region available at some point during the night.}

 \label{fig:observations10}
\end{figure*}

\begin{figure*}
\centering
 \includegraphics[width=0.80\textwidth]{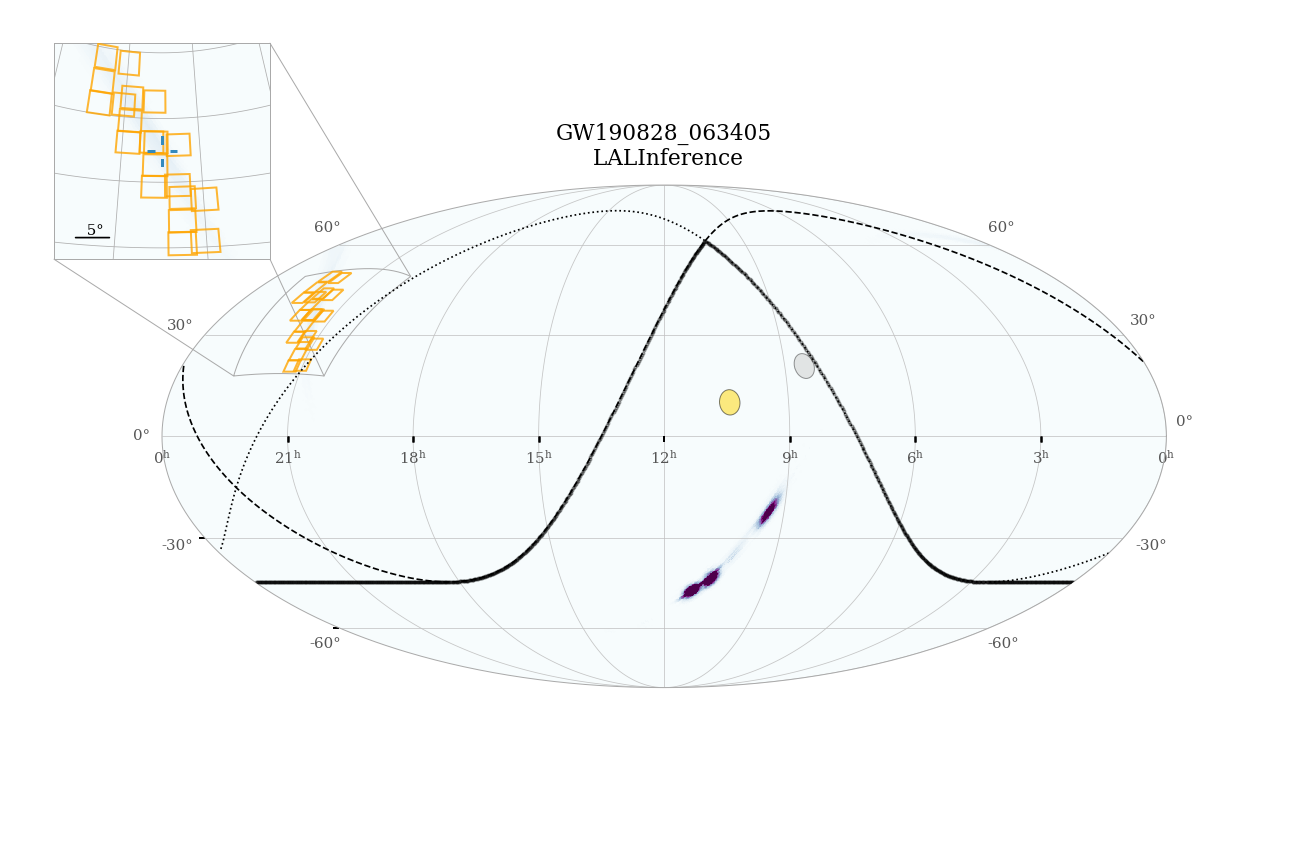}
 \contcaption{DDOTI observations of GW190828$\_$063405. DDOTI fields are shown as orange quadrilaterals and the LIGO/Virgo probability map by blue and purple shading. The Sun and Moon (at the moment when DDOTI began to observe) are indicated by yellow and grey circles respectively. The black dashed line indicates the region of the sky available to DDOTI at the start of the night, the black dotted line indicates the region available to DDOTI at the end of the night, and the black solid line indicates the region available at some point during the night.}

 \label{fig:observations11}
\end{figure*}



\begin{figure*}
\centering
 \includegraphics[width=0.80\textwidth]{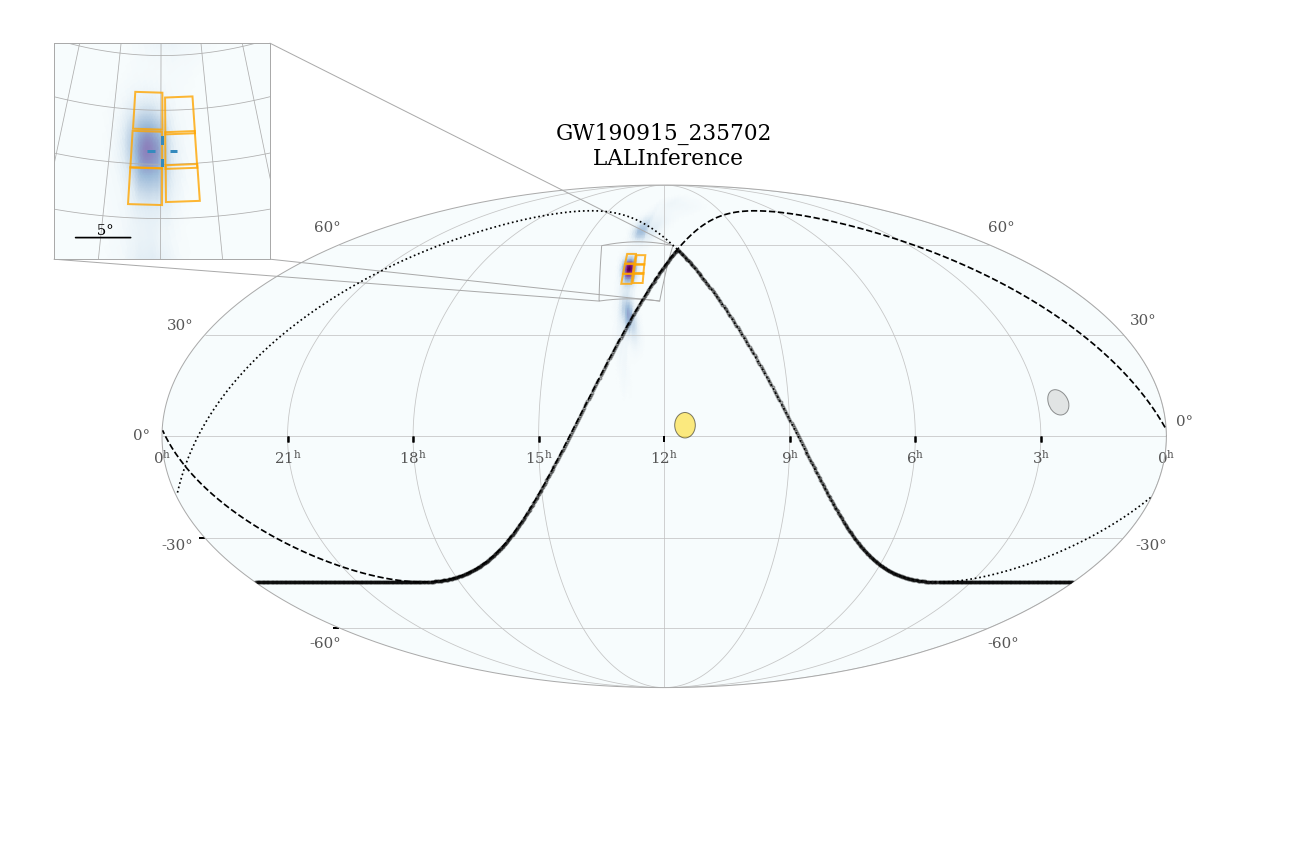}
 \contcaption{DDOTI observations of GW190915$\_$235702. DDOTI fields are shown as orange quadrilaterals and the LIGO/Virgo probability map by blue and purple shading. The Sun and Moon (at the moment when DDOTI began to observe) are indicated by yellow and grey circles respectively. The black dashed line indicates the region of the sky available to DDOTI at the start of the night, the black dotted line indicates the region available to DDOTI at the end of the night, and the black solid line indicates the region available at some point during the night.}

 \label{fig:observations13}
\end{figure*}

%

\begin{figure*}
\centering
 \includegraphics[width=0.80\textwidth]{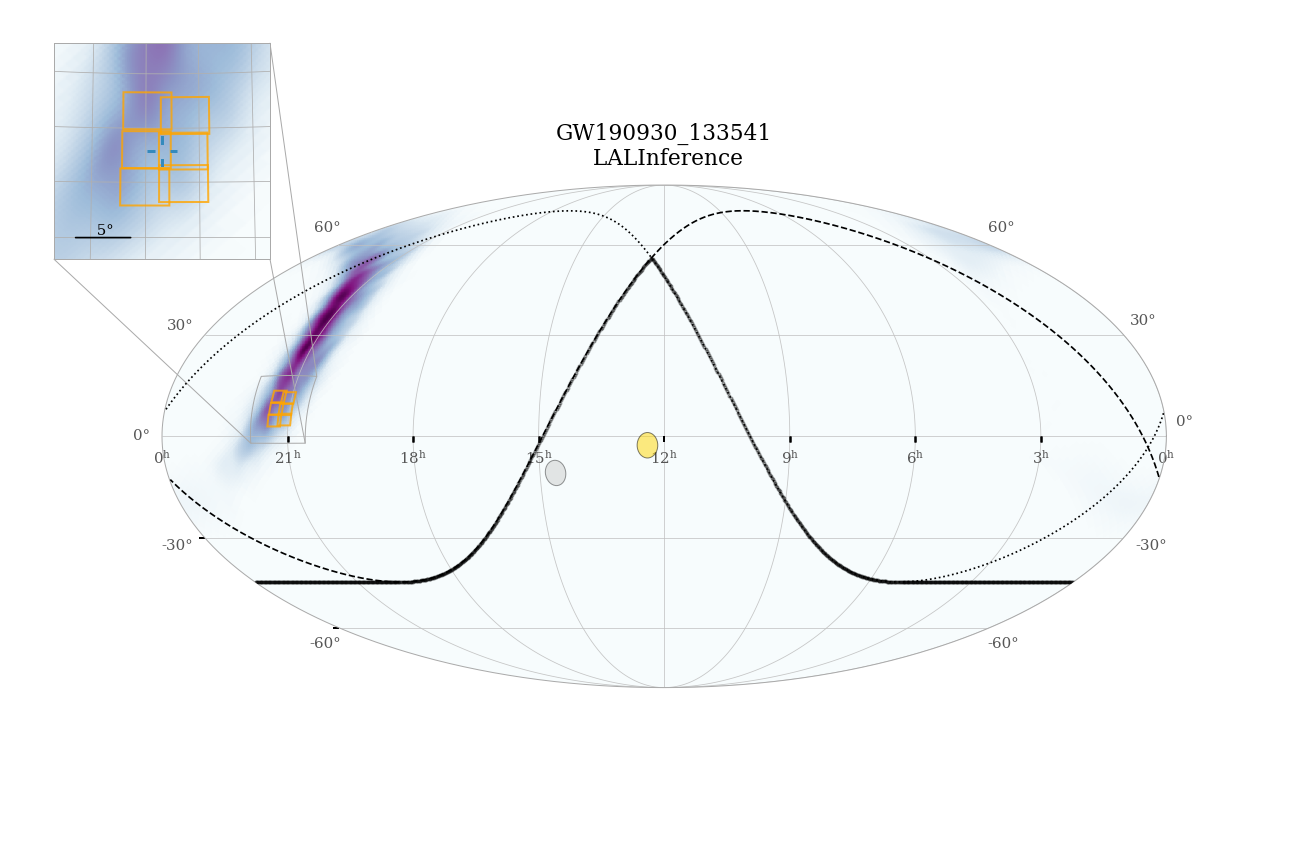}
 \contcaption{DDOTI observations of GW190930$\_$133541. DDOTI fields are shown as orange quadrilaterals and the LIGO/Virgo probability map by blue and purple shading. The Sun and Moon (at the moment when DDOTI began to observe) are indicated by yellow and grey circles respectively. The black dashed line indicates the region of the sky available to DDOTI at the start of the night, the black dotted line indicates the region available to DDOTI at the end of the night, and the black solid line indicates the region available at some point during the night.}

 \label{fig:observations15}
\end{figure*}

\begin{figure*}
\centering
 \includegraphics[width=0.80\textwidth]{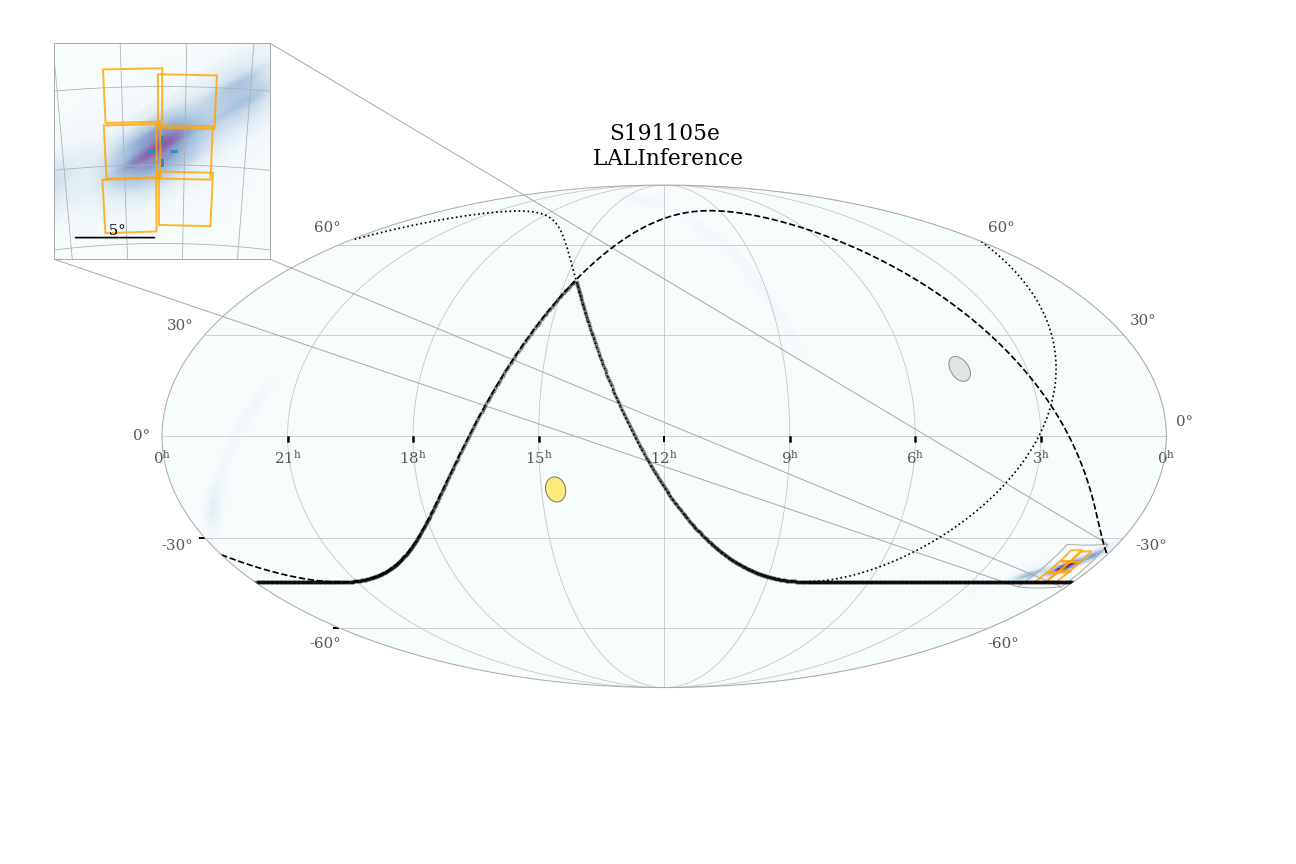}
 \contcaption{DDOTI observations of S191105e. DDOTI fields are shown as orange quadrilaterals and the LIGO/Virgo probability map by blue and purple shading. The Sun and Moon (at the moment when DDOTI began to observe) are indicated by yellow and grey circles respectively. The black dashed line indicates the region of the sky available to DDOTI at the start of the night, the black dotted line indicates the region available to DDOTI at the end of the night, and the black solid line indicates the region available at some point during the night.}

 \label{fig:observations16}
\end{figure*}

\begin{figure*}
\centering
 \includegraphics[width=0.80\textwidth]{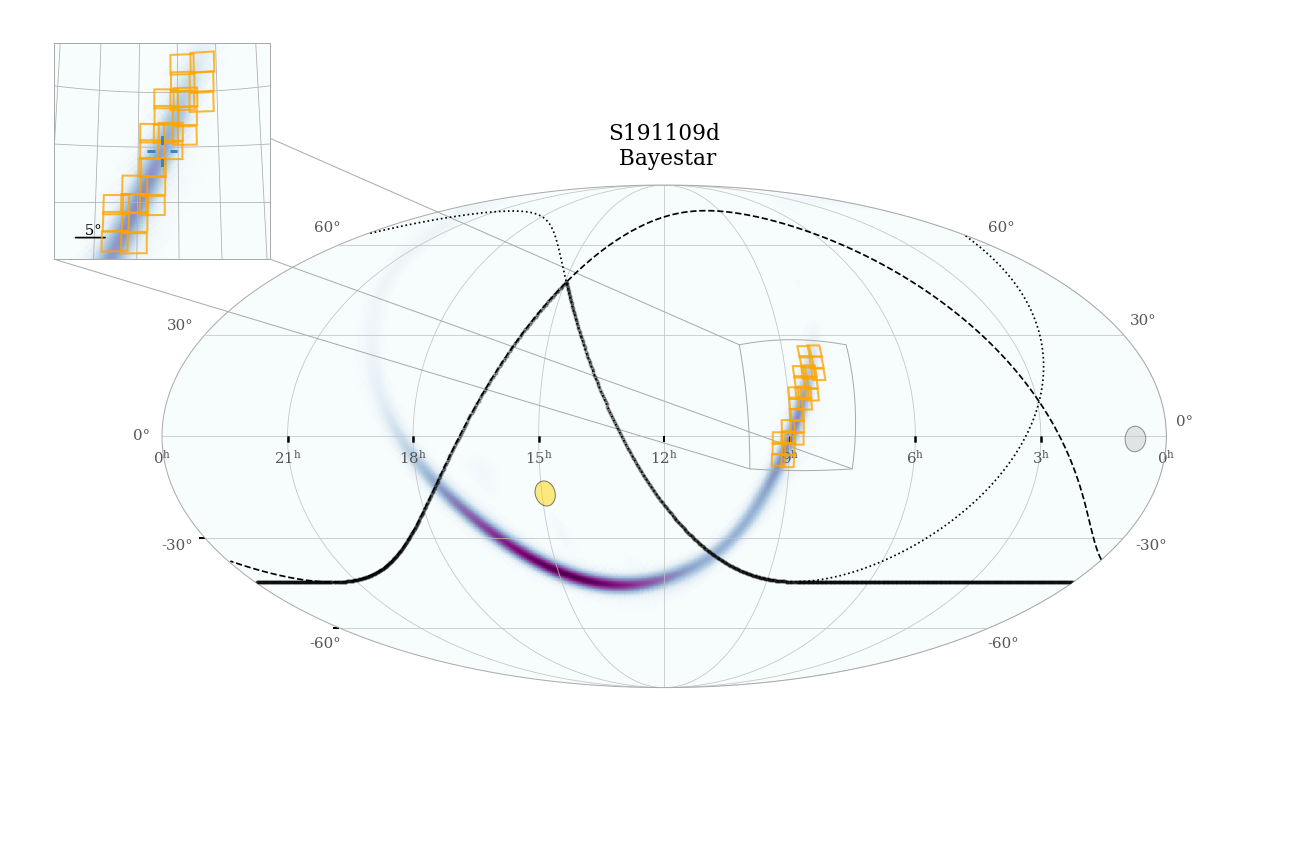}
 \contcaption{DDOTI observations of S191109d. DDOTI fields are shown as orange quadrilaterals and the LIGO/Virgo probability map by blue and purple shading. The Sun and Moon (at the moment when DDOTI began to observe) are indicated by yellow and grey circles respectively. The black dashed line indicates the region of the sky available to DDOTI at the start of the night, the black dotted line indicates the region available to DDOTI at the end of the night, and the black solid line indicates the region available at some point during the night.}

 \label{fig:observations17}
\end{figure*}

\begin{figure*}
\centering
 \includegraphics[width=0.80\textwidth]{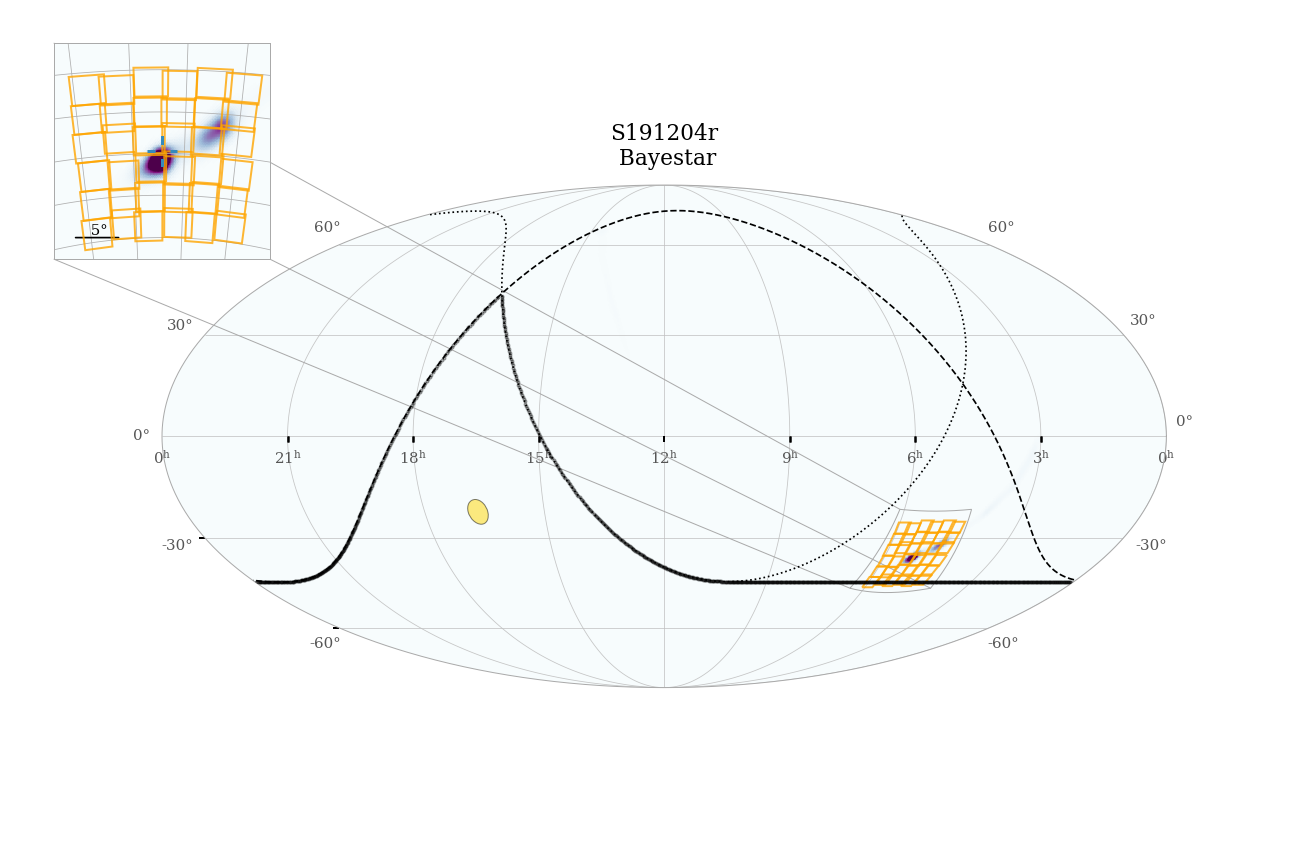}
 \contcaption{DDOTI observations of S191204r. DDOTI fields are shown as orange quadrilaterals and the LIGO/Virgo probability map by blue and purple shading. The Sun and Moon (at the moment when DDOTI began to observe) are indicated by yellow and grey circles respectively. The black dashed line indicates the region of the sky available to DDOTI at the start of the night, the black dotted line indicates the region available to DDOTI at the end of the night, and the black solid line indicates the region available at some point during the night.}

 \label{fig:observations18}
\end{figure*}

\begin{figure*}
\centering
 \includegraphics[width=0.80\textwidth]{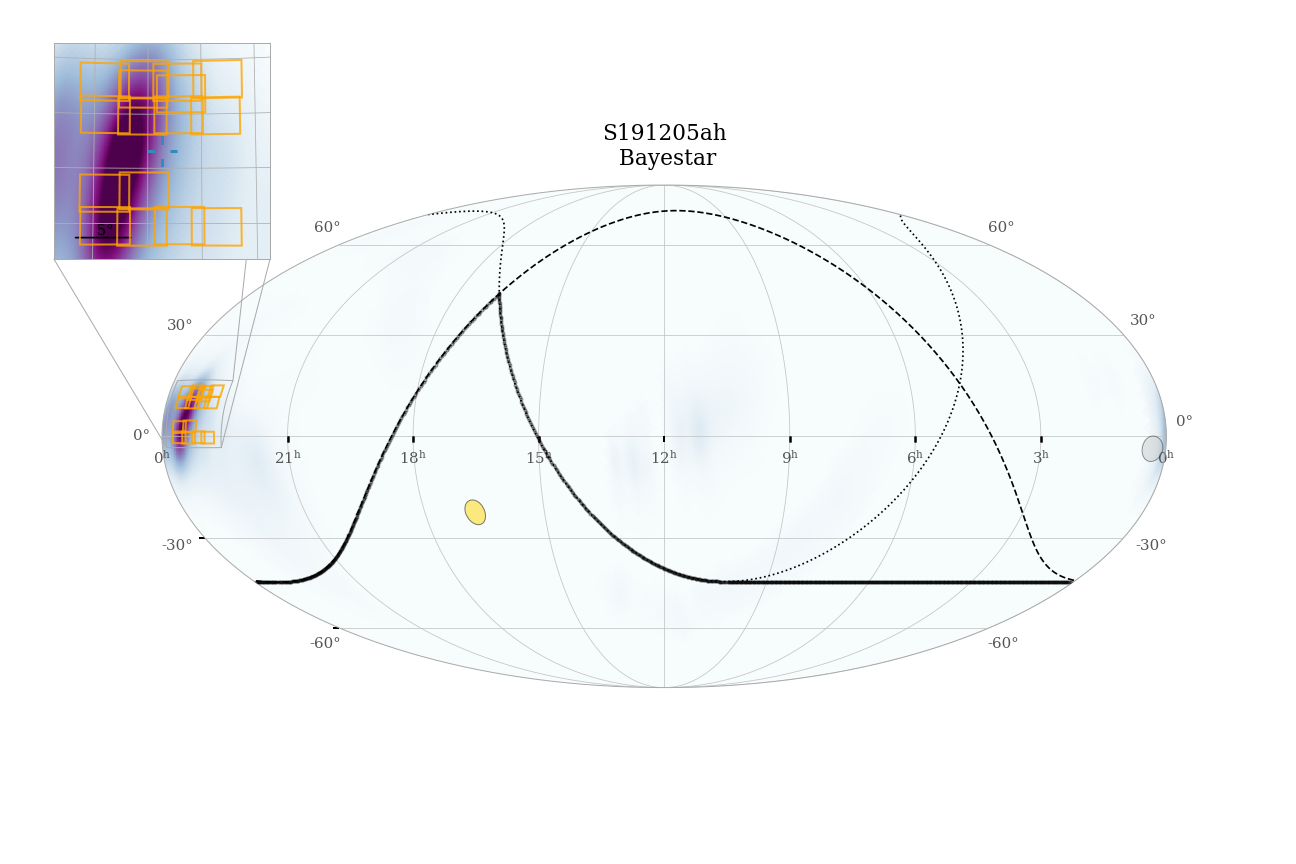}
 \contcaption{DDOTI observations of S191205ah. DDOTI fields are shown as orange quadrilaterals and the LIGO/Virgo probability map by blue and purple shading. The Sun and Moon (at the moment when DDOTI began to observe) are indicated by yellow and grey circles respectively. The black dashed line indicates the region of the sky available to DDOTI at the start of the night, the black dotted line indicates the region available to DDOTI at the end of the night, and the black solid line indicates the region available at some point during the night.}

 \label{fig:observations19}
\end{figure*}

\begin{figure*}
\centering
 \includegraphics[width=0.80\textwidth]{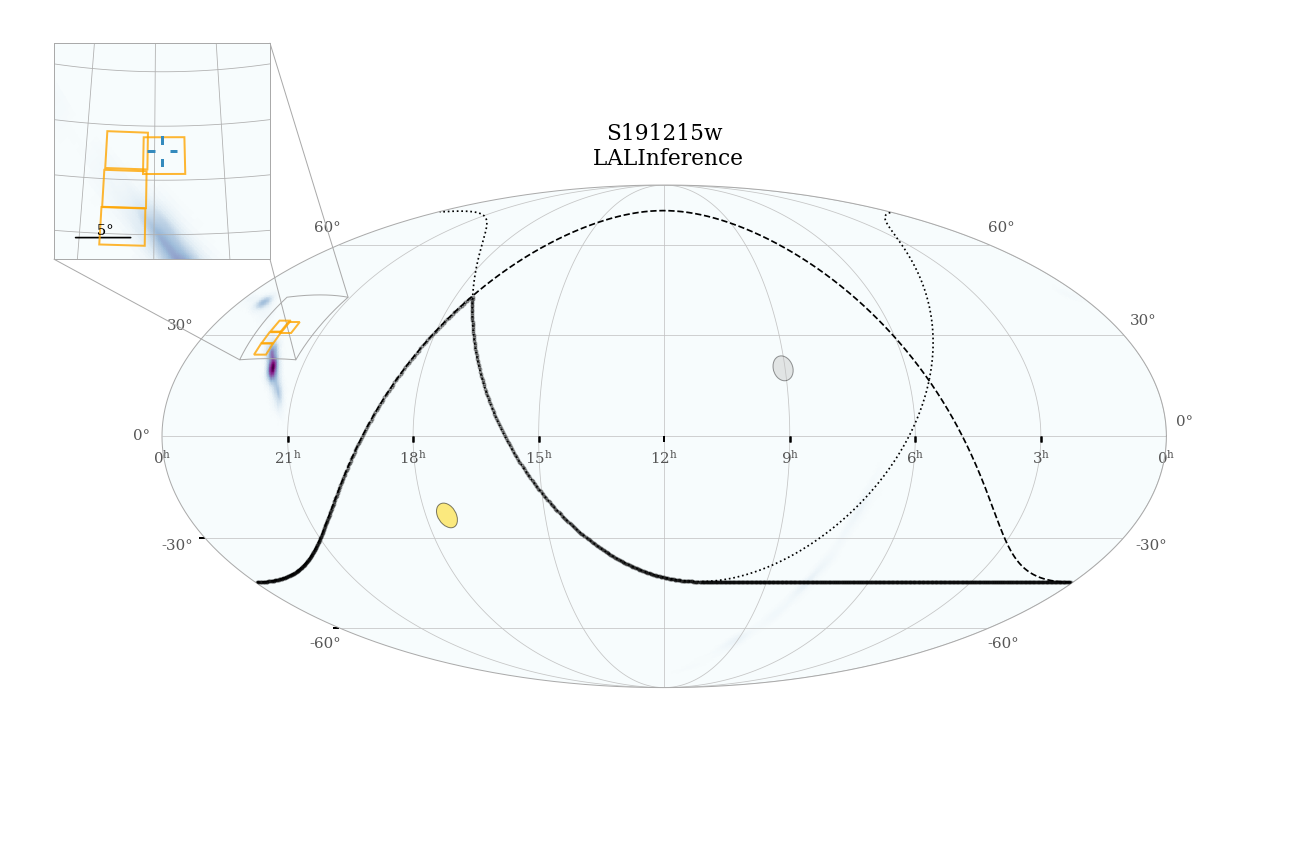}
 \contcaption{DDOTI observations of S191215w. DDOTI fields are shown as orange quadrilaterals and the LIGO/Virgo probability map by blue and purple shading. The Sun and Moon (at the moment when DDOTI began to observe) are indicated by yellow and grey circles respectively. The black dashed line indicates the region of the sky available to DDOTI at the start of the night, the black dotted line indicates the region available to DDOTI at the end of the night, and the black solid line indicates the region available at some point during the night.}

 \label{fig:observations20}
\end{figure*}

\begin{figure*}
\centering
 \includegraphics[width=0.80\textwidth]{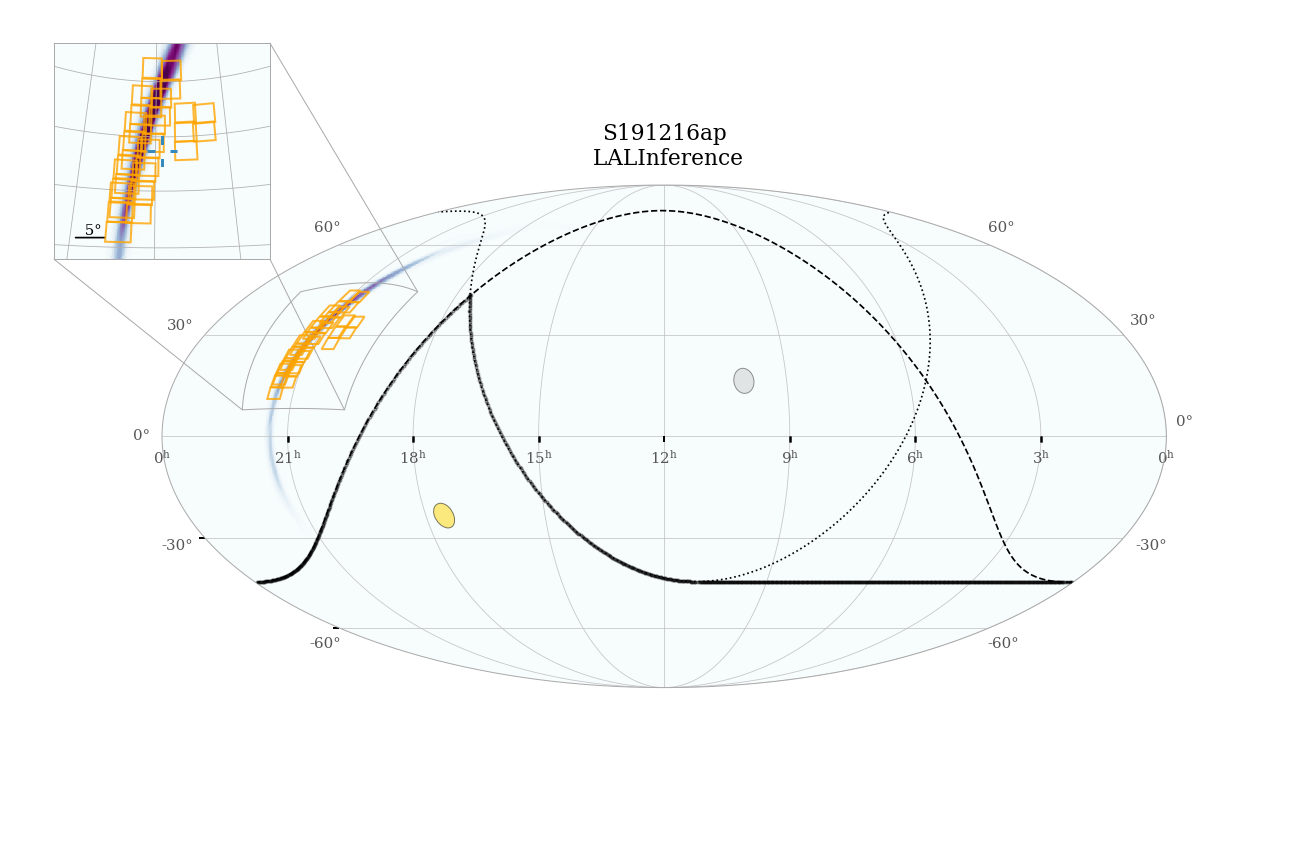}
 \contcaption{DDOTI observations of S191216ap. DDOTI fields are shown as orange quadrilaterals and the LIGO/Virgo probability map by blue and purple shading. The Sun and Moon (at the moment when DDOTI began to observe) are indicated by yellow and grey circles respectively. The black dashed line indicates the region of the sky available to DDOTI at the start of the night, the black dotted line indicates the region available to DDOTI at the end of the night, and the black solid line indicates the region available at some point during the night.}

 \label{fig:observations21}
\end{figure*}

\begin{figure*}
\centering
 \includegraphics[width=0.80\textwidth]{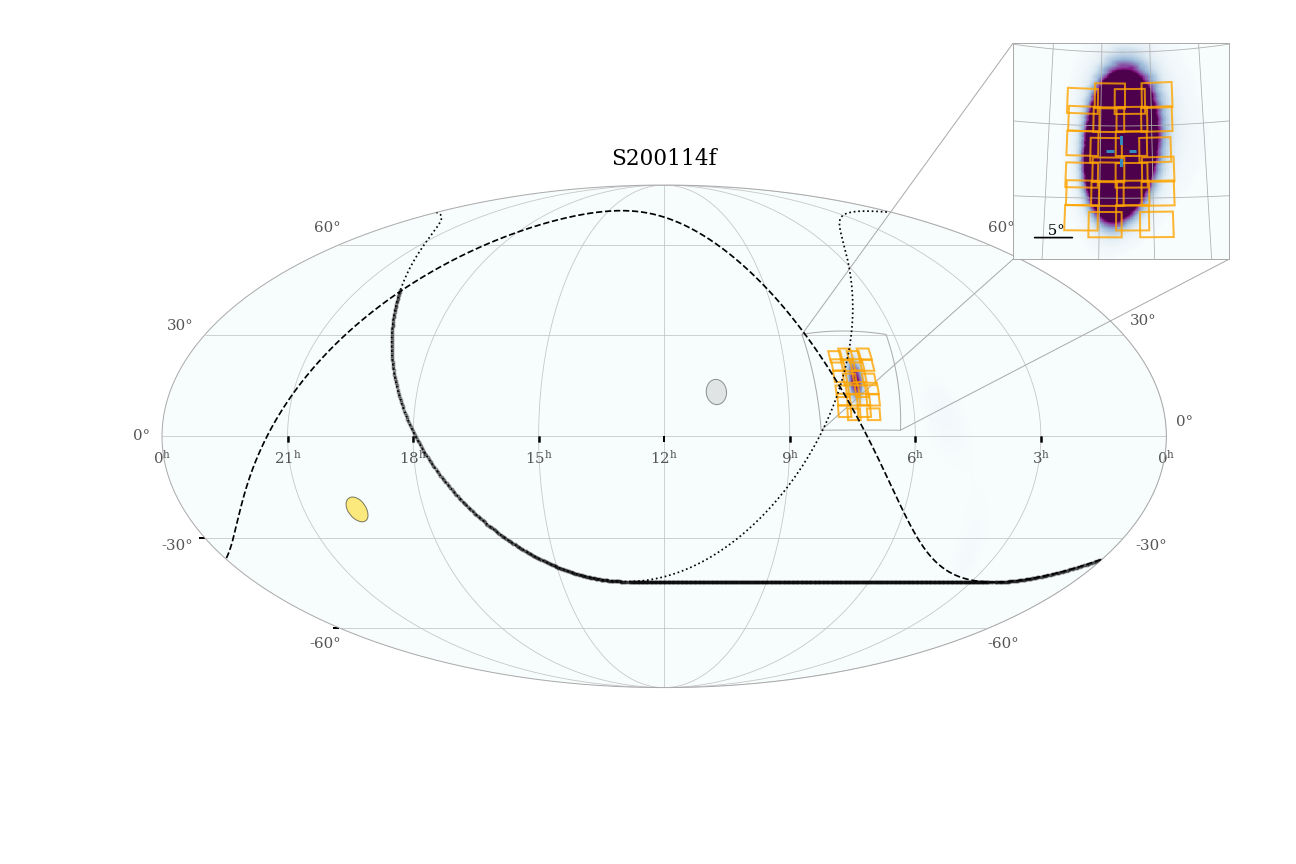}
 \contcaption{DDOTI observations of S200114f. DDOTI fields are shown as orange quadrilaterals and the LIGO/Virgo probability map by blue and purple shading. The Sun and Moon (at the moment when DDOTI began to observe) are indicated by yellow and grey circles respectively. The black dashed line indicates the region of the sky available to DDOTI at the start of the night, the black dotted line indicates the region available to DDOTI at the end of the night, and the black solid line indicates the region available at some point during the night.}

 \label{fig:observations22}
\end{figure*}

\begin{figure*}
\centering
 \includegraphics[width=0.80\textwidth]{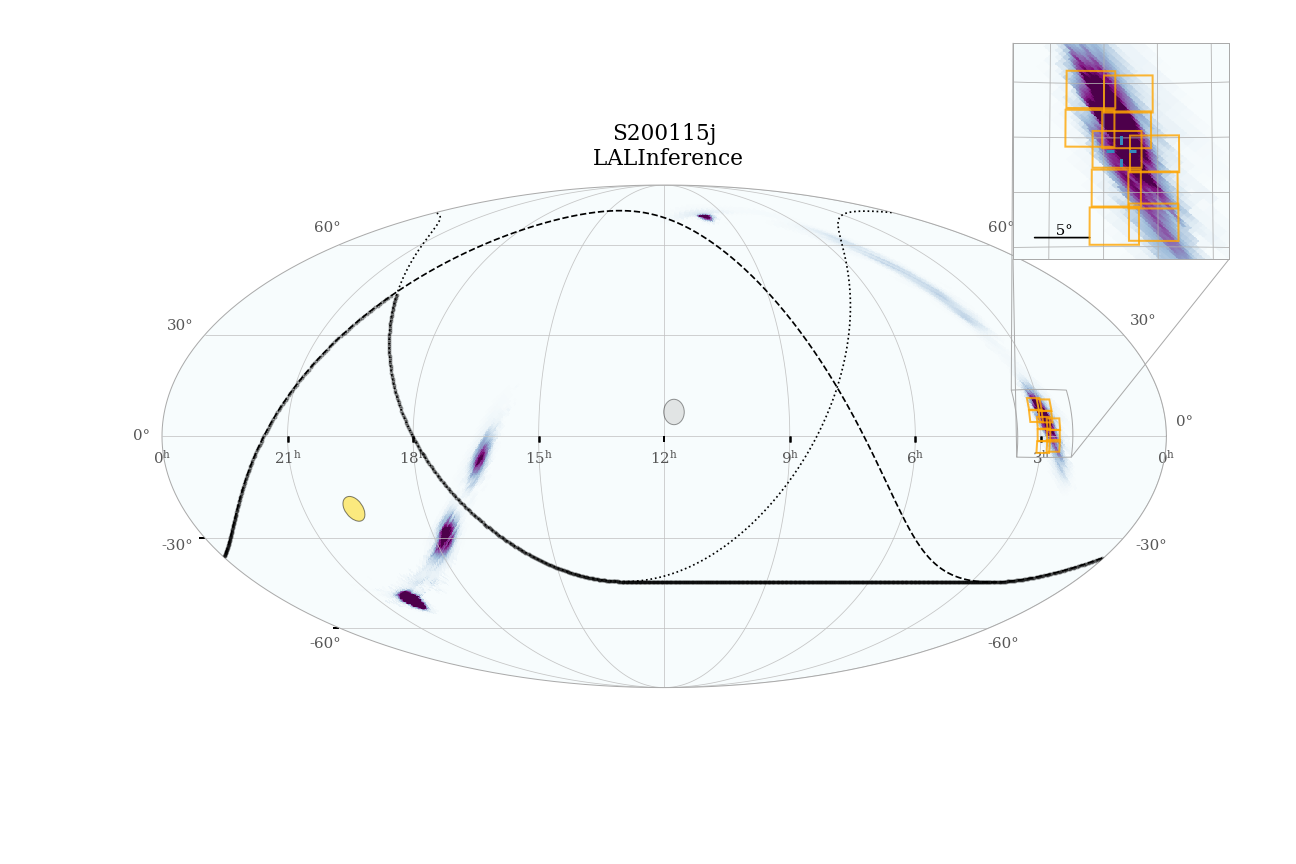}
 \contcaption{DDOTI observations of S200115j. DDOTI fields are shown as orange quadrilaterals and the LIGO/Virgo probability map by blue and purple shading. The Sun and Moon (at the moment when DDOTI began to observe) are indicated by yellow and grey circles respectively. The black dashed line indicates the region of the sky available to DDOTI at the start of the night, the black dotted line indicates the region available to DDOTI at the end of the night, and the black solid line indicates the region available at some point during the night.}

 \label{fig:observations23}
\end{figure*}

\begin{figure*}
\centering
 \includegraphics[width=0.80\textwidth]{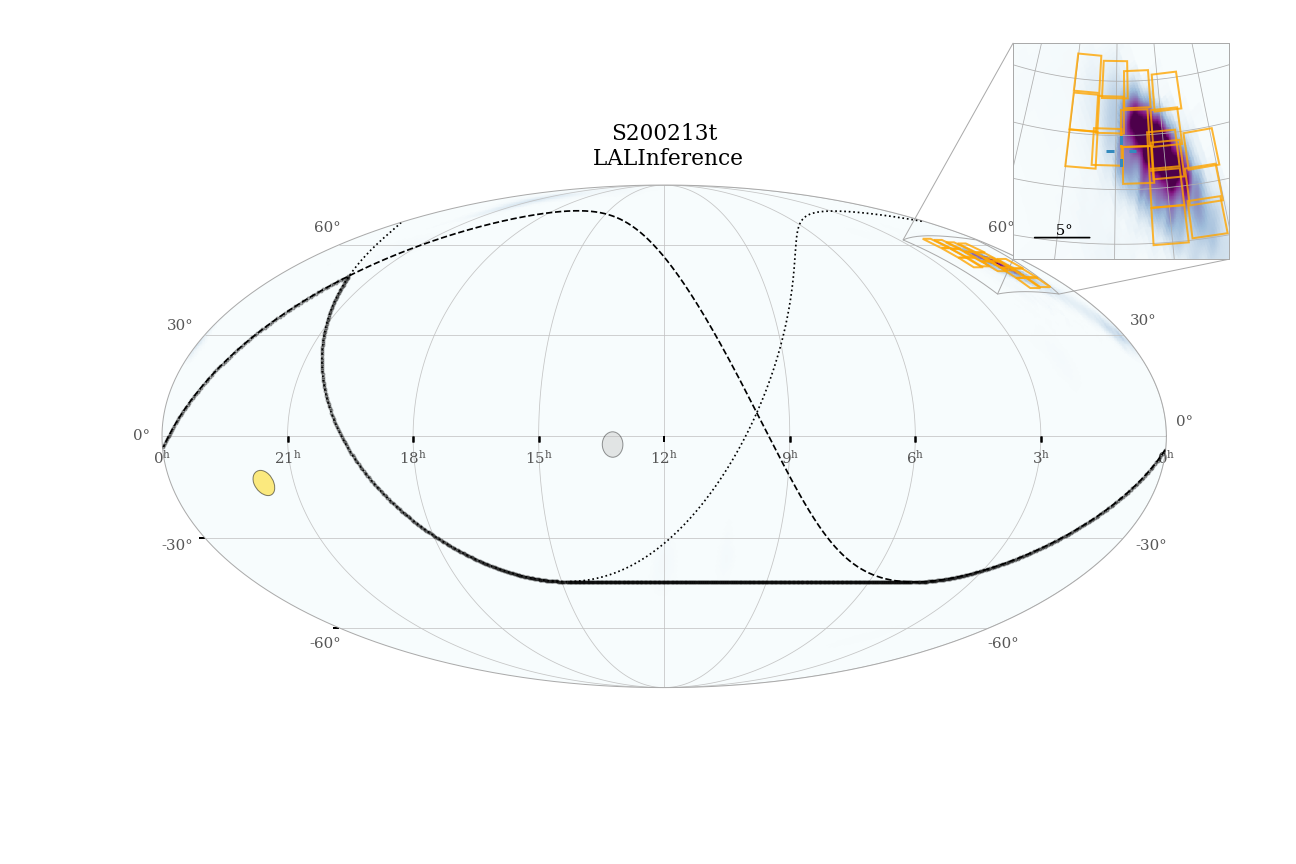}
 \contcaption{DDOTI observations of S200213t. DDOTI fields are shown as orange quadrilaterals and the LIGO/Virgo probability map by blue and purple shading. The Sun and Moon (at the moment when DDOTI began to observe) are indicated by yellow and grey circles respectively. The black dashed line indicates the region of the sky available to DDOTI at the start of the night, the black dotted line indicates the region available to DDOTI at the end of the night, and the black solid line indicates the region available at some point during the night.}

 \label{fig:observations24}
\end{figure*}

\begin{figure*}
\centering
 \includegraphics[width=0.80\textwidth]{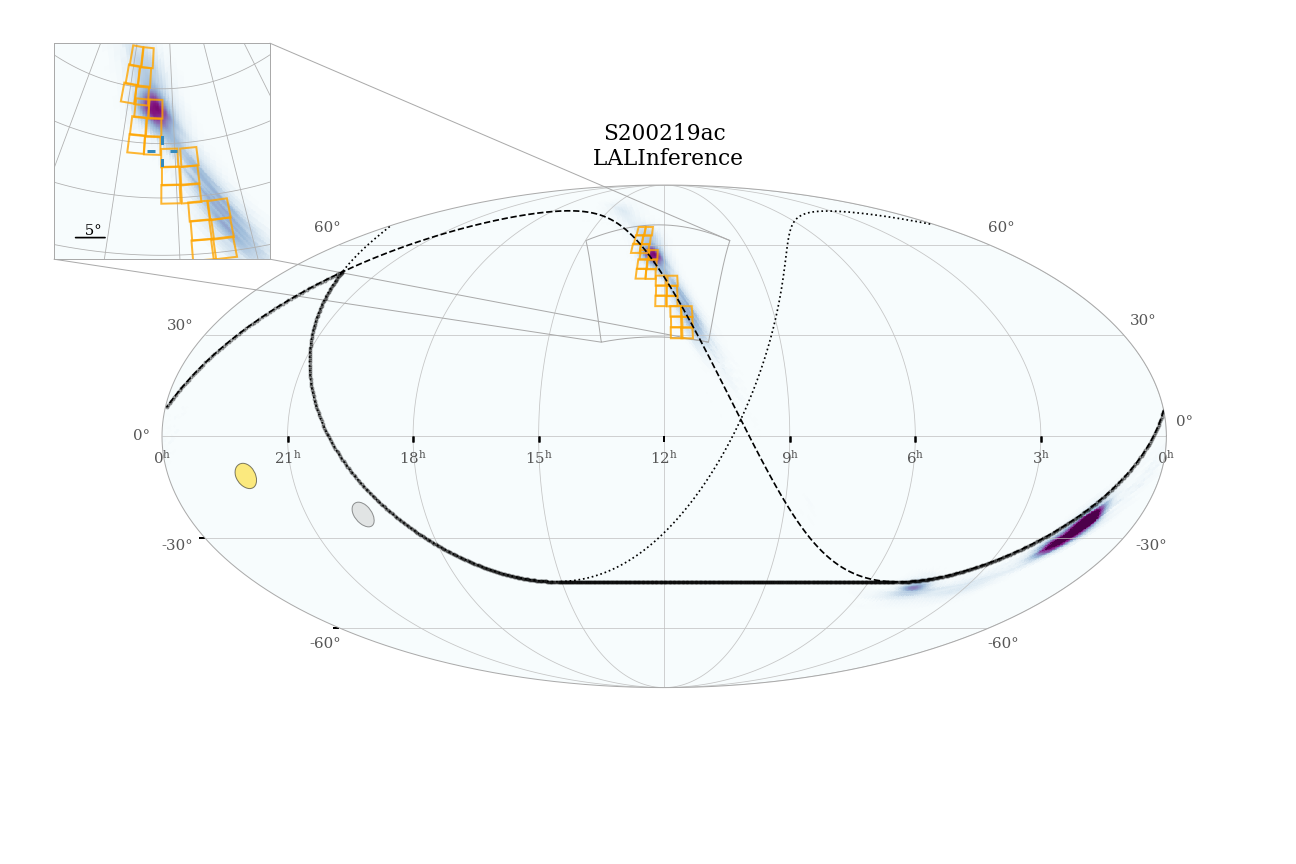}
 \contcaption{DDOTI observations of S200219ac. DDOTI fields are shown as orange quadrilaterals and the LIGO/Virgo probability map by blue and purple shading. The Sun and Moon (at the moment when DDOTI began to observe) are indicated by yellow and grey circles respectively. The black dashed line indicates the region of the sky available to DDOTI at the start of the night, the black dotted line indicates the region available to DDOTI at the end of the night, and the black solid line indicates the region available at some point during the night.}

 \label{fig:observations25}
\end{figure*}

\begin{figure*}
\centering
 \includegraphics[width=0.80\textwidth]{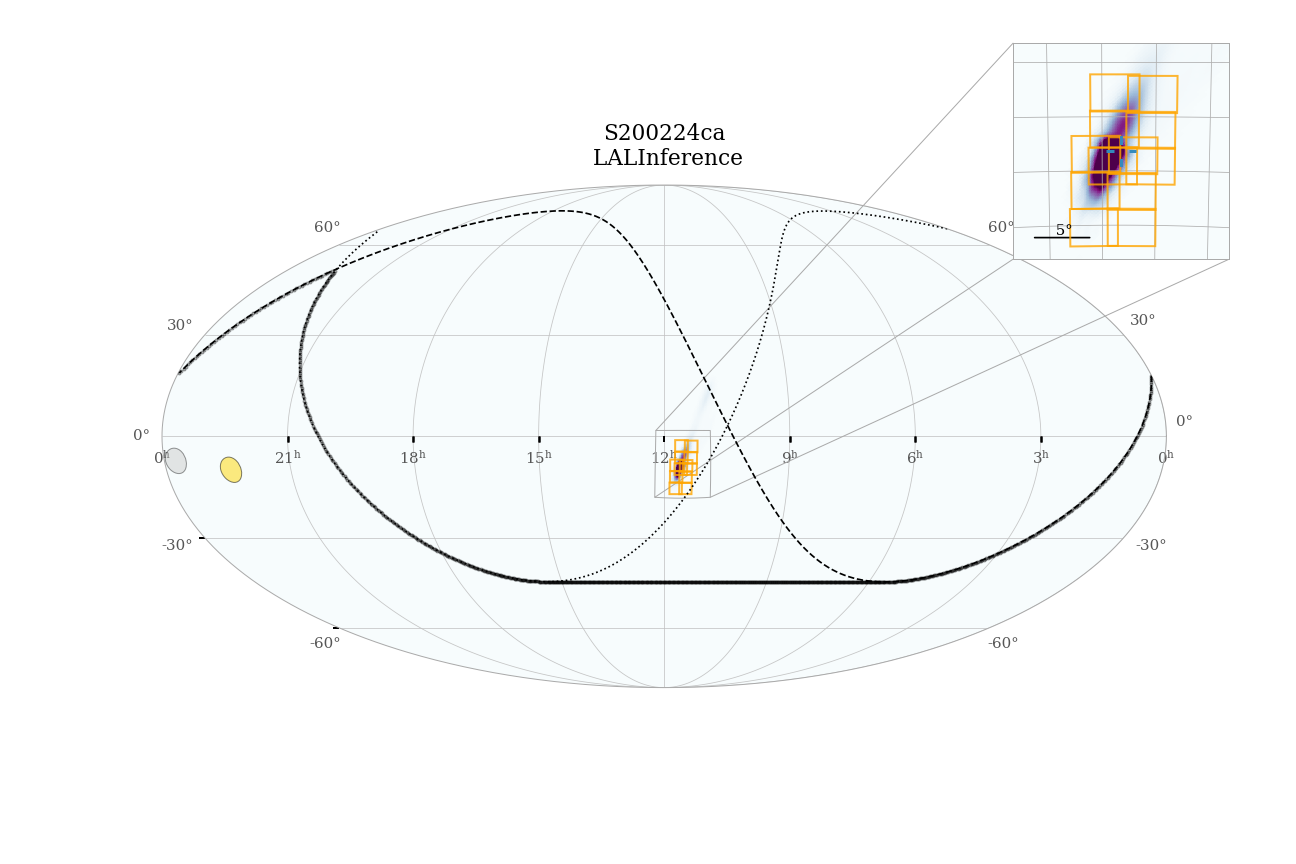}
 \contcaption{DDOTI observations of S200224ca. DDOTI fields are shown as orange quadrilaterals and the LIGO/Virgo probability map by blue and purple shading. The Sun and Moon (at the moment when DDOTI began to observe) are indicated by yellow and grey circles respectively. The black dashed line indicates the region of the sky available to DDOTI at the start of the night, the black dotted line indicates the region available to DDOTI at the end of the night, and the black solid line indicates the region available at some point during the night.}

 \label{fig:observations26}
\end{figure*}

\begin{figure*}
\centering
 \includegraphics[width=0.80\textwidth]{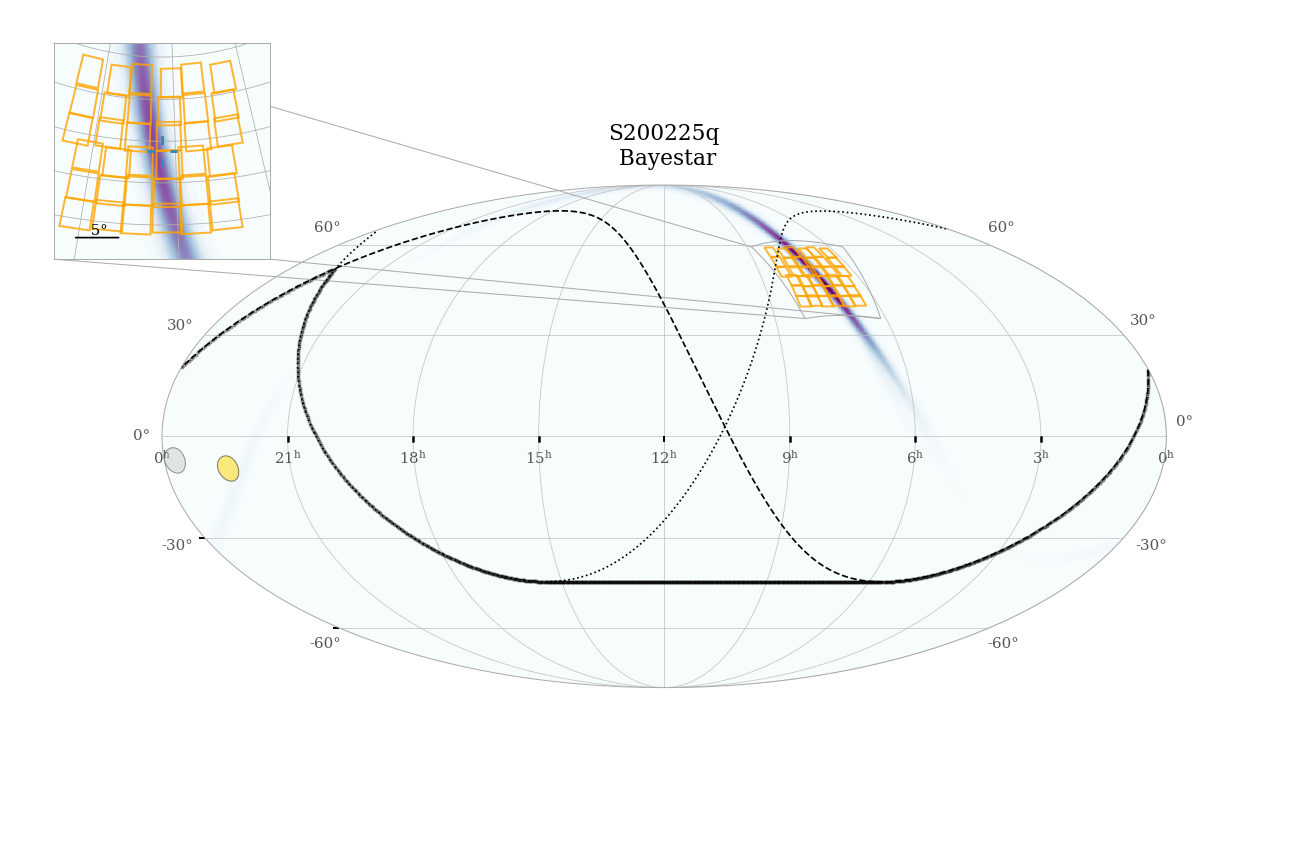}
 \contcaption{DDOTI observations of S200225q. DDOTI fields are shown as orange quadrilaterals and the LIGO/Virgo probability map by blue and purple shading. The Sun and Moon (at the moment when DDOTI began to observe) are indicated by yellow and grey circles respectively. The black dashed line indicates the region of the sky available to DDOTI at the start of the night, the black dotted line indicates the region available to DDOTI at the end of the night, and the black solid line indicates the region available at some point during the night.}

 \label{fig:observations27}
\end{figure*}

\begin{figure*}
\centering
 \includegraphics[width=0.80\textwidth]{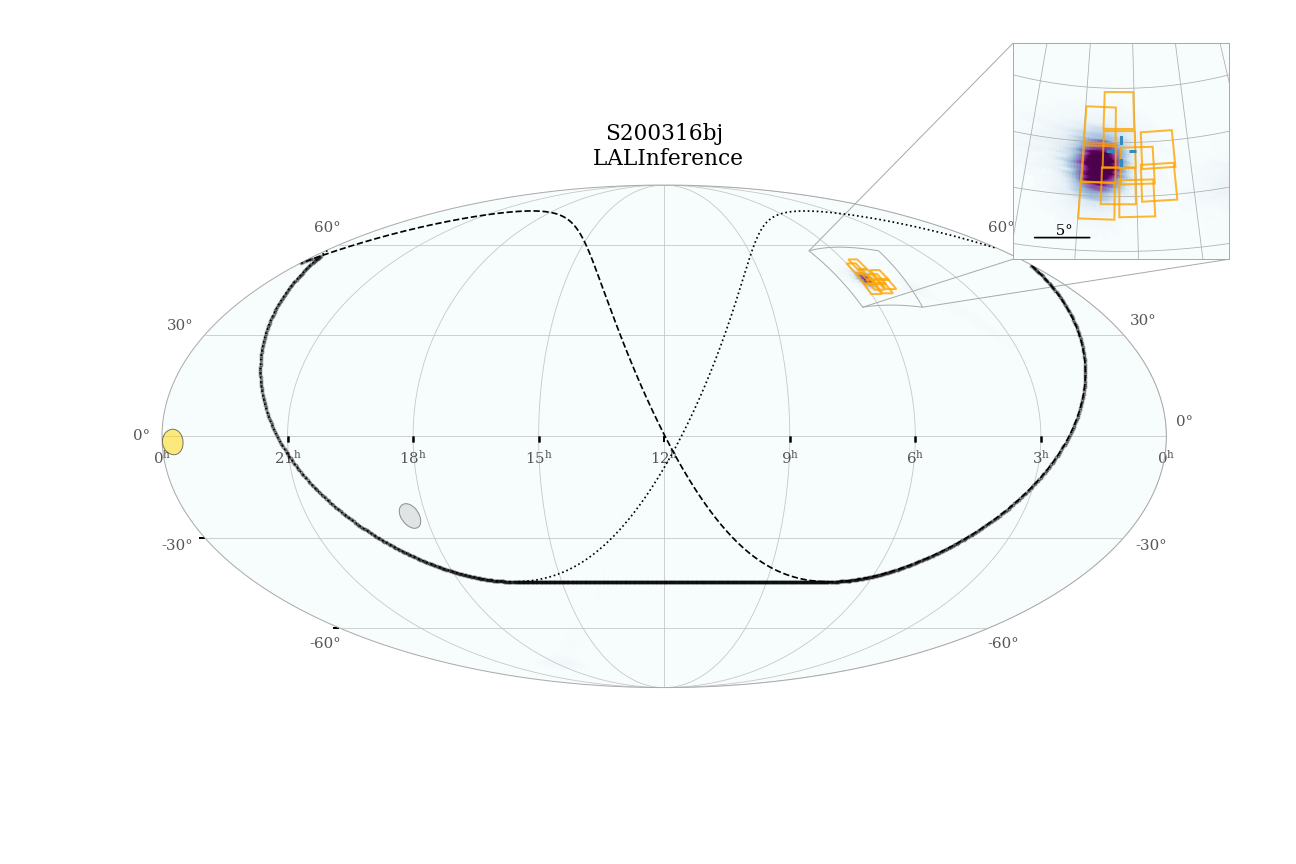}
 \contcaption{DDOTI observations of S200316bj. DDOTI fields are shown as orange quadrilaterals and the LIGO/Virgo probability map by blue and purple shading. The Sun and Moon (at the moment when DDOTI began to observe) are indicated by yellow and grey circles respectively. The black dashed line indicates the region of the sky available to DDOTI at the start of the night, the black dotted line indicates the region available to DDOTI at the end of the night, and the black solid line indicates the region available at some point during the night.}

 \label{fig:observations28}
\end{figure*}

\section*{Research Data Policy}
The data underlying this article will be shared on reasonable request to the corresponding author.\\

\section*{ACKNOWLEDGEMENTS}
We thank Noel Klingler for the review of the article and his comments.
We thank the staff of the Observatorio Astron\'omico Nacional.
DDOTI is funded by CONACyT (LN 260369, LN 271117, and 277901), the University of Maryland (NNX17AK54G), and the Universidad Nacional Autónoma de México (CIC and DGAPA/PAPIIT IG100414, IT102715, AG100317, and IN109418, AG100820, IN105921) and is operated and maintained by the Observatorio Astron\'omico Nacional and the Instituto de Astronom\'ia of the Universidad Nacional Aut\'onoma de México. 
RLB acknowledges support from the DGAPA-UNAM postdoctoral fellowship.





\begin{thebibliography}{99}
\bibitem[\protect\citeauthoryear{Abbott et al.}{2016a}]{Abbott2016a} Abbott B.~P., Abbott R., Abbott T.~D., Abernathy M.~R., Acernese F., Ackley K., Adams C., et al., 2016a, ApJL, 818, L22. doi:10.3847/2041-8205/818/2/L22

\bibitem[\protect\citeauthoryear{Abbott et al.}{2016b}]{Abbott2016b} Abbott B.~P., et al., 2016b, PhRvL, 116, 061102

\bibitem[\protect\citeauthoryear{Abbott et al.}{2017a}]{Abbott2017a} Abbott B.~P., et al., 2017a, PhRvL, 119, 161101

\bibitem[\protect\citeauthoryear{Abbott et al.}{2017b}]{Abbott2017b} Abbott B.~P., et al., 2017b, ApJL, 848, L12

\bibitem[\protect\citeauthoryear{Abbott, et al.}{2019}]{Abbott2019} Abbott B.~P., et al., 2019, PhRvD, 100, 064064

\bibitem[Abbott et al.(2020a)]{Abbott2020a} Abbott, B.~P., Abbott, R., Abbott, T.~D., et al.\ 2020a Living Reviews in Relativity, 23, 3. doi:10.1007/s41114-020-00026-9

\bibitem[\protect\citeauthoryear{Abbott et al.}{2020b}]{LVC190814} Abbott R., Abbott T.~D., Abraham S., Acernese F., Ackley K., Adams C., Adhikari R.~X., et al., 2020b, ApJL, 896, L44. doi:10.3847/2041-8213/ab960f

\bibitem[\protect\citeauthoryear{Abbott et al.}{2020c}]{Abbott2020c} Abbott R., Abbott T.~D., Abraham S., Acernese F., Ackley K., Adams A., Adams C., et al., 2020c, arXiv, arXiv:2010.14527

\bibitem[\protect\citeauthoryear{Abbott et al.}{2020d}]{LIGOpop} Abbott R., Abbott T.~D., Abraham S., Acernese F., Ackley K., et al., 2020d, arXiv, arXiv:2010.14533


\bibitem[\protect\citeauthoryear{Ackley et al.}{2020}]{Ackley2020} Ackley K., Amati L., Barbieri C., Bauer F.~E., Benetti S., Bernardini M.~G., Bhirombhakdi K., et al., 2020, A\&A, 643, A113. doi:10.1051/0004-6361/202037669

\bibitem[\protect\citeauthoryear{Ahn et al.}{2012}]{Ahn2012} Ahn C.~P., Alexandroff R., Allende Prieto C., Anderson S.~F., Anderton T., Andrews B.~H., Aubourg {\'E}., et al., 2012, ApJS, 203, 21. doi:10.1088/0067-0049/203/2/21

\bibitem[\protect\citeauthoryear{Alexander et al.}{2021}]{Alexander2021} Alexander K.~D., Schroeder G., Paterson K., Fong W., Cowperthwaite P., Gomez S., Margalit B., et al., 2021, arXiv, arXiv:2102.08957

\bibitem[\protect\citeauthoryear{Andreoni et al.}{2017}]{Andreoni2017} Andreoni I., Ackley K., Cooke J., Acharyya A., Allison J.~R., Anderson G.~E., Ashley M.~C.~B., et al., 2017, PASA, 34, e069. doi:10.1017/pasa.2017.65


\bibitem[Andreoni et al.(2020)]{Andreoni2020} Andreoni, I., Goldstein, D.~A., Kasliwal, M.~M., et al.\ 2020, \apj, 890, 131

\bibitem[\protect\citeauthoryear{Andreoni et al.}{2021}]{Andreoni2021} Andreoni I., Coughlin M.~W., Kool E.~C., Kasliwal M.~M., Kumar H., Bhalerao V., Sagu{\'e}s Carracedo A., et al., 2021, arXiv, arXiv:2104.06352

\bibitem[\protect\citeauthoryear{Antier et al.}{2020}]{Antier2020} Antier S., Agayeva S., Almualla M., Awiphan S., Baransky A., Barynova K., Beradze S., et al., 2020, MNRAS, 497, 5518. doi:10.1093/mnras/staa1846

\bibitem[\protect\citeauthoryear{Arcavi et al.}{2017}]{Arcavi2017} Arcavi I., Hosseinzadeh G., Howell D.~A., McCully C., Poznanski D., Kasen D., Barnes J., et al., 2017, Natur, 551, 64. doi:10.1038/nature24291

\bibitem[\protect\citeauthoryear{Ascenzi et al.}{2021}]{Ascenzi2021} Ascenzi S., Oganesyan G., Branchesi M., Ciolfi R., 2021, JPlPh, 87, 845870102. doi:10.1017/S0022377820001646

\bibitem[\protect\citeauthoryear{Barbieri et al.}{2019}]{Barbieri2019} Barbieri C., Salafia O.~S., Colpi M., Ghirlanda G., Perego A., Colombo A., 2019, ApJL, 887, L35. doi:10.3847/2041-8213/ab5c1e

\bibitem[\protect\citeauthoryear{Barbieri et al.}{2020}]{Barbieri2020} Barbieri C., Salafia O.~S., Colpi M., Ghirlanda G., Perego A., 2020, arXiv, arXiv:2002.09395

\bibitem[\protect\citeauthoryear{Bartos et al.}{2017}]{Bartos2017} Bartos I., Kocsis B., Haiman Z., M{\'a}rka S., 2017, ApJ, 835, 165. doi:10.3847/1538-4357/835/2/165

\bibitem[\protect\citeauthoryear{Becerra et al.}{2019}]{Becerra2019} Becerra R.~L., Dichiara S., Watson A.~M., Troja E., Fraija N., Klotz A., Butler N.~R., et al., 2019, ApJ, 881, 12. doi:10.3847/1538-4357/ab275b

\bibitem[\protect\citeauthoryear{Becerra et al.}{2021}]{Becerra2021} Becerra R.~L., De Colle F., Cant{\'o} J., Lizano S., Gonz{\'a}lez R.~F., Granot J., Klotz A., et al., 2021, ApJ, 908, 39. doi:10.3847/1538-4357/abcd3a

\bibitem[\protect\citeauthoryear{Beniamini \& van der Horst}{2017}]{BeniaminivanderHorst} Beniamini P., van der Horst A.~J., 2017, MNRAS, 472, 3161. doi:10.1093/mnras/stx2203

\bibitem[\protect\citeauthoryear{Bennett et al.}{2014}]{Bennett2014} Bennett C.~L., Larson D., Weiland J.~L., Hinshaw G., 2014, ApJ, 794, 135. doi:10.1088/0004-637X/794/2/135

\bibitem[\protect\citeauthoryear{Bernuzzi et al.}{2020}]{Bernuzzi2020} Bernuzzi S., Breschi M., Daszuta B., Endrizzi A., Logoteta D., Nedora V., Perego A., et al., 2020, MNRAS, 497, 1488. doi:10.1093/mnras/staa1860


\bibitem[\protect\citeauthoryear{Bertin \& Arnouts}{1996}]{Bertin1996} Bertin E., Arnouts S., 1996, A\&AS, 117, 393

\bibitem[\protect\citeauthoryear{Bertin}{2010}]{Bertin2010} Bertin E., 2010, Astrophysics Source Code Library, ascl:1010.068

\bibitem[\protect\citeauthoryear{Bruni et al.}{2021}]{Bruni2021} Bruni G., O'Connor B., Matsumoto T., Troja E., Piran T., Piro L., Ricci R., 2021, MNRAS.tmp. doi:10.1093/mnrasl/slab046

\bibitem[\protect\citeauthoryear{Chang et al.}{2021}]{Skymapper} Chang S.-W., Onken C.~A., Wolf C., Luvaul L., M{\"o}ller A., Scalzo R., Schmidt B.~P., et al., 2021, arXiv, arXiv:2102.07353

\bibitem[\protect\citeauthoryear{Chase et al.}{2021}]{Chase2021} Chase E.~A., O'Connor B., Fryer C.~L., Troja E., Korobkin O., Wollaeger R.~T., Ristic M., et al., 2021, arXiv, arXiv:2105.12268

\bibitem[\protect\citeauthoryear{Connaughton et al.}{2016}]{Connaughton2016} Connaughton V., Burns E., Goldstein A., Blackburn L., Briggs M.~S., Zhang B.-B., Camp J., et al., 2016, ApJL, 826, L6. doi:10.3847/2041-8205/826/1/L6

\bibitem[\protect\citeauthoryear{Coulter et al.}{2017}]{Coulter2017} Coulter D.~A., et al., 2017, Science, 358, 1556

\bibitem[\protect\citeauthoryear{Cromartie et al.}{2020}]{Cromartie2020} Cromartie H.~T., Fonseca E., Ransom S.~M., Demorest P.~B., Arzoumanian Z., Blumer H., Brook P.~R., et al., 2020, NatAs, 4, 72. doi:10.1038/s41550-019-0880-2

\bibitem[\protect\citeauthoryear{Darbha \& Kasen}{2020}]{Darbha2020} Darbha S., Kasen D., 2020, ApJ, 897, 150. doi:10.3847/1538-4357/ab9a34

\bibitem[\protect\citeauthoryear{D{\'\i}az et al.}{2017}]{Diaz2017} D{\'\i}az M.~C., Macri L.~M., Garcia Lambas D., Mendes de Oliveira C., Nilo Castell{\'o}n J.~L., Ribeiro T., S{\'a}nchez B., et al., 2017, ApJL, 848, L29. doi:10.3847/2041-8213/aa9060

\bibitem[\protect\citeauthoryear{Dichiara et al.}{2019}]{2019GCN.25352....1D} Dichiara S., Pereyra M., Watson A.~M., Butler N., Becerra R.~L., Gonzalez D., Kutyrev A., et al., 2019, GCN, 25352

\bibitem[\protect\citeauthoryear{Dichiara \& Wolfram}{2020}]{2020GCN.26752....1D} Dichiara S., Wolfram T., 2020, GCN, 26752, 1

\bibitem[\protect\citeauthoryear{Dichiara, et al.}{2020a}]{2020GCN.27212....1D} Dichiara S., et al., 2020a, GCN, 27212, 1

\bibitem[\protect\citeauthoryear{Dichiara et al.}{2020b}]{Dichiara2020} Dichiara S., Troja E., O'Connor B., Marshall F.~E., Beniamini P., Cannizzo J.~K., Lien A.~Y., et al., 2020b, MNRAS, 492, 5011. doi:10.1093/mnras/staa124

\bibitem[\protect\citeauthoryear{Dichiara et al.}{2021}]{Dichiara2021} Dichiara S., Troja E., Beniamini P., O'Connor B., Moss M., Lien A.~Y., Ricci R., et al., 2021, ApJL, 911, L28. doi:10.3847/2041-8213/abf562

\bibitem[\protect\citeauthoryear{Dobie et al.}{2019}]{Dobie2019} Dobie D., Stewart A., Murphy T., Lenc E., Wang Z., Kaplan D.~L., Andreoni I., et al., 2019, ApJL, 887, L13. doi:10.3847/2041-8213/ab59db

\bibitem[\protect\citeauthoryear{Estell{\'e}s et al.}{2021}]{Estelle2021} Estell{\'e}s H., Husa S., Colleoni M., Mateu-Lucena M., de Lluc Planas M., Garc{\'\i}a-Quir{\'o}s C., Keitel D., et al., 2021, arXiv, arXiv:2105.06360


\bibitem[\protect\citeauthoryear{Evans et al.}{2016}]{Evans2016} Evans P.~A., Kennea J.~A., Palmer D.~M., Bilicki M., Osborne J.~P., O'Brien P.~T., Tanvir N.~R., et al., 2016, MNRAS, 462, 1591. doi:10.1093/mnras/stw1746


\bibitem[\protect\citeauthoryear{Evans et al.}{2017}]{Evans2017} Evans P.~A., Cenko S.~B., Kennea J.~A., Emery S.~W.~K., Kuin N.~P.~M., Korobkin O., Wollaeger R.~T., et al., 2017, Sci, 358, 1565. doi:10.1126/science.aap9580

\bibitem[\protect\citeauthoryear{Fern{\'a}ndez, Foucart, \& Lippuner}{2020}]{Fernandez2020} Fern{\'a}ndez R., Foucart F., Lippuner J., 2020, MNRAS, 497, 3221. doi:10.1093/mnras/staa2209

\bibitem[\protect\citeauthoryear{Fontes et al.}{2020}]{Fontes2020} Fontes C.~J., Fryer C.~L., Hungerford A.~L., Wollaeger R.~T., Korobkin O., 2020, MNRAS, 493, 4143. doi:10.1093/mnras/staa485

\bibitem[\protect\citeauthoryear{Fujibayashi et al.}{2020}]{Fujibashi2020} Fujibayashi S., Wanajo S., Kiuchi K., Kyutoku K., Sekiguchi Y., Shibata M., 2020, ApJ, 901, 122. doi:10.3847/1538-4357/abafc2

\bibitem[\protect\citeauthoryear{Gao \& Fan}{2006}]{Gao2006} Gao W.-H., Fan Y.-Z., 2006, ChJAA, 6, 513. doi:10.1088/1009-9271/6/5/01

\bibitem[\protect\citeauthoryear{Gao et al.}{2017}]{Gao2017} Gao H., Zhang B., L{\"u} H.-J., Li Y., 2017, ApJ, 837, 50. doi:10.3847/1538-4357/aa5be3

\bibitem[\protect\citeauthoryear{Gehrels et al.}{2016}]{Gehrels2016} Gehrels N., Cannizzo J.~K., Kanner J., Kasliwal M.~M., Nissanke S., Singer L.~P., 2016, ApJ, 820, 136. doi:10.3847/0004-637X/820/2/136

\bibitem[\protect\citeauthoryear{Goldstein, et al.}{2017}]{Goldstein2017} Goldstein A., et al., 2017, ApJL, 848, L14

\bibitem[\protect\citeauthoryear{Golkhou et al.}{2018}]{Golkhou2018} Golkhou V.~Z., Butler N.~R., Strausbaugh R., Troja E., Kutyrev A., Lee W.~H., Rom{\'a}n-Z{\'u}{\~n}iga C.~G., et al., 2018, ApJ, 857, 81. doi:10.3847/1538-4357/aab665


\bibitem[\protect\citeauthoryear{Gompertz et al.}{2020}]{Gompertz2020} Gompertz B.~P., Cutter R., Steeghs D., Galloway D.~K., Lyman J., Ulaczyk K., Dyer M.~J., et al., 2020, MNRAS, 497, 726. doi:10.1093/mnras/staa1845

\bibitem[\protect\citeauthoryear{Graham et al.}{2020}]{Graham2020} Graham M.~J., Ford K.~E.~S., McKernan B., Ross N.~P., Stern D., Burdge K., Coughlin M., et al., 2020, PhRvL, 124, 251102. doi:10.1103/PhysRevLett.124.251102

\bibitem[\protect\citeauthoryear{Granot, De Colle, \& Ramirez-Ruiz}{2018}]{Granot2018} Granot J., De Colle F., Ramirez-Ruiz E., 2018, MNRAS, 481, 2711. doi:10.1093/mnras/sty2454


\bibitem[\protect\citeauthoryear{Hallinan et al.}{2017}]{Hallinan2017} Hallinan G., Corsi A., Mooley K.~P., Hotokezaka K., Nakar E., Kasliwal M.~M., Kaplan D.~L., et al., 2017, Sci, 358, 1579. doi:10.1126/science.aap9855


\bibitem[\protect\citeauthoryear{Henden et al.}{2018}]{Henden2018} Henden A.~A., Levine S., Terrell D., Welch D.~L., Munari U., Kloppenborg B.~K., 2018, AAS, 232, 223.06

\bibitem[\protect\citeauthoryear{Hu et al.}{2017}]{Hu2017} Hu L., Wu X., Andreoni I., Ashley M.~C.~B., Cooke J., Cui X., Du F., et al., 2017, SciBu, 62, 1433. doi:10.1016/j.scib.2017.10.006

\bibitem[\protect\citeauthoryear{Jiang et al.}{2019}]{Jiang2019} Jiang Y.-F., Blaes O., Stone J.~M., Davis S.~W., 2019, ApJ, 885, 144. doi:10.3847/1538-4357/ab4a00

\bibitem[\protect\citeauthoryear{Kawaguchi et al.}{2015}]{Kawaguchi2015} Kawaguchi K., Kyutoku K., Nakano H., Okawa H., Shibata M., Taniguchi K., 2015, PhRvD, 92, 024014. doi:10.1103/PhysRevD.92.024014

\bibitem[\protect\citeauthoryear{Kawaguchi et al.}{2020}]{Kawaguchi2020} Kawaguchi K., Fujibayashi S., Shibata M., Tanaka M., Wanajo S., 2020, arXiv, arXiv:2012.14711


\bibitem[\protect\citeauthoryear{Klingler et al.}{2021}]{Klinger2021} Klingler N.~J., Lien A., Oates S.~R., Kennea J.~A., Evans P.~A., Tohuvavohu A., Zhang B., et al., 2021, ApJ, 907, 97. doi:10.3847/1538-4357/abd2c3

\bibitem[\protect\citeauthoryear{Korobkin et al.}{2021}]{Korobkin2021} Korobkin O., Wollaeger R.~T., Fryer C.~L., Hungerford A.~L., Rosswog S., Fontes C.~J., Mumpower M.~R., et al., 2021, ApJ, 910, 116. doi:10.3847/1538-4357/abe1b5

\bibitem[\protect\citeauthoryear{Kr{\"u}ger \& Foucart}{2020}]{Kruger2020} Kr{\"u}ger C.~J., Foucart F., 2020, PhRvD, 101, 103002. doi:10.1103/PhysRevD.101.103002

\bibitem[Kumar \& Zhang(2015)]{Kumar2015} Kumar, P. \& Zhang, B.\ 2015, \physrep, 561, 1. doi:10.1016/j.physrep.2014.09.008

\bibitem[\protect\citeauthoryear{Lang, et al.}{2010}]{Lang2010} Lang D., Hogg D.~W., Mierle K., Blanton M., Roweis S., 2010, AJ, 139, 1782

\bibitem[\protect\citeauthoryear{Lazzati et al.}{2018}]{Lazzati2018} Lazzati D., Perna R., Morsony B.~J., Lopez-Camara D., Cantiello M., Ciolfi R., Giacomazzo B., et al., 2018, PhRvL, 120, 241103. doi:10.1103/PhysRevLett.120.241103

\bibitem[\protect\citeauthoryear{Liu, Gao, \& Zhang}{2020}]{Li2020} Liu L.-D., Gao H., Zhang B., 2020, ApJ, 890, 102. doi:10.3847/1538-4357/ab6b24

\bibitem[\protect\citeauthoryear{LIGO Scientific Collaboration, et al.}{2015}]{LVC15} LIGO Scientific Collaboration, et al., 2015, CQGra, 32, 074001

\bibitem[\protect\citeauthoryear{Lipunov et al.}{2017}]{Lipunov2017} Lipunov V.~M., Gorbovskoy E., Kornilov V.~G., . Tyurina N., Balanutsa P., Kuznetsov A., Vlasenko D., et al., 2017, ApJL, 850, L1. doi:10.3847/2041-8213/aa92c0

\bibitem[\protect\citeauthoryear{L{\"u} et al.}{2015}]{Lu2015} L{\"u} H.-J., Zhang B., Lei W.-H., Li Y., Lasky P.~D., 2015, ApJ, 805, 89. doi:10.1088/0004-637X/805/2/89

\bibitem[\protect\citeauthoryear{McKernan et al.}{2020}]{McKernan2020} McKernan B., Ford K.~E.~S., O'Shaugnessy R., Wysocki D., 2020, MNRAS, 494, 1203. doi:10.1093/mnras/staa740

\bibitem[\protect\citeauthoryear{Miller et al.}{2019}]{Miller2019} Miller M.~C., Lamb F.~K., Dittmann A.~J., Bogdanov S., Arzoumanian Z., Gendreau K.~C., Guillot S., et al., 2019, ApJL, 887, L24. doi:10.3847/2041-8213/ab50c5

\bibitem[Monet et al.(2003)]{Monet2003} Monet, D.~G., Levine, S.~E., Canzian, B., et al.\ 2003, \aj, 125, 984. doi:10.1086/345888

\bibitem[\protect\citeauthoryear{Morgan et al.}{2020}]{Morgan2020} Morgan R., Soares-Santos M., Annis J., Herner K., Garcia A., Palmese A., Drlica-Wagner A., et al., 2020, ApJ, 901, 83. doi:10.3847/1538-4357/abafaa

\bibitem[\protect\citeauthoryear{Most et al.}{2020}]{Most2020} Most E.~R., Papenfort L.~J., Tootle S., Rezzolla L., 2020, arXiv, arXiv:2012.03896

\bibitem[\protect\citeauthoryear{Nakar}{2020}]{Nakar2020} Nakar E., 2020, PhR, 886, 1. doi:10.1016/j.physrep.2020.08.008

\bibitem[\protect\citeauthoryear{O'Connor, Beniamini, \& Kouveliotou}{2020}]{Oconnor2020} O'Connor B., Beniamini P., Kouveliotou C., 2020, MNRAS, 495, 4782. doi:10.1093/mnras/staa1433

\bibitem[\protect\citeauthoryear{O'Connor et al.}{2021}]{Oconnor2021} O'Connor B., Troja E., Dichiara S., Chase E.~A., Ryan G., Cenko S.~B., Fryer C.~L., et al., 2021, MNRAS, 502, 1279. doi:10.1093/mnras/stab132

\bibitem[\protect\citeauthoryear{Page et al.}{2020}]{Page2020} Page K.~L., Evans P.~A., Tohuvavohu A., Kennea J.~A., Klingler N.~J., Cenko S.~B., Oates S.~R., et al., 2020, MNRAS, 499, 3459. doi:10.1093/mnras/staa3032


\bibitem[\protect\citeauthoryear{Pereyra, et al.}{2019}]{2019GCN.25562....1P} Pereyra M., et al., 2019, GCN, 25562, 1

\bibitem[\protect\citeauthoryear{Pereyra, et al.}{2020}]{2020GCN.27402....1P} Pereyra M., et al., 2020, GCN, 27402, 1

\bibitem[\protect\citeauthoryear{Perna, Lazzati, \& Giacomazzo}{2016}]{Perna2016} Perna R., Lazzati D., Giacomazzo B., 2016, ApJL, 821, L18. doi:10.3847/2041-8205/821/1/L18

\bibitem[\protect\citeauthoryear{Pian et al.}{2017}]{Pian2017} Pian E., D'Avanzo P., Benetti S., Branchesi M., Brocato E., Campana S., Cappellaro E., et al., 2017, Natur, 551, 67. doi:10.1038/nature24298

\bibitem[\protect\citeauthoryear{Ricci et al.}{2021}]{Ricci2021} Ricci R., Troja E., Bruni G., Matsumoto T., Piro L., O'Connor B., Piran T., et al., 2021, MNRAS, 500, 1708. doi:10.1093/mnras/staa3241

\bibitem[\protect\citeauthoryear{Riley et al.}{2019}]{Riley2019} Riley T.~E., Watts A.~L., Bogdanov S., Ray P.~S., Ludlam R.~M., Guillot S., Arzoumanian Z., et al., 2019, ApJL, 887, L21. doi:10.3847/2041-8213/ab481c

\bibitem[\protect\citeauthoryear{Rowlinson et al.}{2013}]{Rowlinson2013} Rowlinson A., O'Brien P.~T., Metzger B.~D., Tanvir N.~R., Levan A.~J., 2013, MNRAS, 430, 1061. doi:10.1093/mnras/sts683

\bibitem[\protect\citeauthoryear{Ryan et al.}{2020}]{Ryan2020} Ryan G., van Eerten H., Piro L., Troja E., 2020, ApJ, 896, 166. doi:10.3847/1538-4357/ab93cf

\bibitem[\protect\citeauthoryear{Sagu{\'e}s Carracedo et al.}{2021}]{Sagues2021} Sagu{\'e}s Carracedo A., Bulla M., Feindt U., Goobar A., 2021, MNRAS, 504, 1294. doi:10.1093/mnras/stab872

\bibitem[\protect\citeauthoryear{Santana, Barniol Duran, \& Kumar}{2014}]{Santana14} Santana R., Barniol Duran R., Kumar P., 2014, ApJ, 785, 29. doi:10.1088/0004-637X/785/1/29

\bibitem[\protect\citeauthoryear{Sarin \& Lasky}{2020}]{Sarin2020} Sarin N., Lasky P.~D., 2020, arXiv, arXiv:2012.08172

\bibitem[\protect\citeauthoryear{Savchenko et al.}{2016}]{Savchenko2016} Savchenko V., Ferrigno C., Mereghetti S., Natalucci L., Bazzano A., Bozzo E., Brandt S., et al., 2016, ApJL, 820, L36. doi:10.3847/2041-8205/820/2/L36

\bibitem[\protect\citeauthoryear{Savchenko et al.}{2017}]{Savchenko2017} Savchenko V., Ferrigno C., Kuulkers E., Bazzano A., Bozzo E., Brandt S., Chenevez J., et al., 2017, ApJL, 848, L15. doi:10.3847/2041-8213/aa8f94

\bibitem[\protect\citeauthoryear{Schutz}{2011}]{Schutz2011} Schutz B.~F., 2011, CQGra, 28, 125023. doi:10.1088/0264-9381/28/12/125023

\bibitem[\protect\citeauthoryear{Singer \& Price}{2016}]{Singer2016} Singer L.~P., Price L.~R., 2016, PhRvD, 93, 024013. doi:10.1103/PhysRevD.93.024013

\bibitem[\protect\citeauthoryear{Soares-Santos et al.}{2017}]{Soares2017} Soares-Santos M., Holz D.~E., Annis J., Chornock R., Herner K., Berger E., Brout D., et al., 2017, ApJL, 848, L16. doi:10.3847/2041-8213/aa9059


\bibitem[\protect\citeauthoryear{Tanvir et al.}{2017}]{Tanvir2017} Tanvir N.~R., Levan A.~J., Gonz{\'a}lez-Fern{\'a}ndez C., Korobkin O., Mandel I., Rosswog S., Hjorth J., et al., 2017, ApJL, 848, L27. doi:10.3847/2041-8213/aa90b6

\bibitem[\protect\citeauthoryear{Thakur et al.}{2020}]{Thakur2020} Thakur A.~L., Dichiara S., Troja E., Chase E.~A., S{\'a}nchez-Ram{\'\i}rez R., Piro L., Fryer C.~L., et al., 2020, MNRAS, 499, 3868. doi:10.1093/mnras/staa2798

\bibitem[Tonry et al.(2012)]{Tonry2012} Tonry, J.~L., Stubbs, C.~W., Lykke, K.~R., et al.\ 2012, \apj, 750, 99. doi:10.1088/0004-637X/750/2/99

\bibitem[\protect\citeauthoryear{Troja et al.}{2016}]{Troja2016} Troja E., Read A.~M., Tiengo A., Salvaterra R., 2016, ApJL, 822, L8. doi:10.3847/2041-8205/822/1/L8

\bibitem[\protect\citeauthoryear{Troja et al.}{2017}]{Troja2017} Troja E., Piro L., van Eerten H., Wollaeger R.~T., Im M., Fox O.~D., Butler N.~R., et al., 2017, Natur, 551, 71. doi:10.1038/nature24290

\bibitem[\protect\citeauthoryear{Troja et al.}{2019}]{Troja19} Troja E., van Eerten H., Ryan G., Ricci R., Burgess J.~M., Wieringa M.~H., Piro L., et al., 2019, MNRAS, 489, 1919. doi:10.1093/mnras/stz2248


\bibitem[\protect\citeauthoryear{Typel et al.}{2010}]{Typel2010} Typel S., R{\"o}pke G., Kl{\"a}hn T., Blaschke D., Wolter H.~H., 2010, PhRvC, 81, 015803. doi:10.1103/PhysRevC.81.015803

\bibitem[\protect\citeauthoryear{Urrutia et al.}{2021}]{Urrutia2021} Urrutia G., De Colle F., Murguia-Berthier A., Ramirez-Ruiz E., 2021, MNRAS, 503, 4363. doi:10.1093/mnras/stab723

\bibitem[\protect\citeauthoryear{Utsumi et al.}{2017}]{Utsumi2017} Utsumi Y., Tanaka M., Tominaga N., Yoshida M., Barway S., Nagayama T., Zenko T., et al., 2017, PASJ, 69, 101. doi:10.1093/pasj/psx118

\bibitem[\protect\citeauthoryear{Valenti et al.}{2017}]{Valenti2017} Valenti S., Sand D.~J., Yang S., Cappellaro E., Tartaglia L., Corsi A., Jha S.~W., et al., 2017, ApJL, 848, L24. doi:10.3847/2041-8213/aa8edf

\bibitem[\protect\citeauthoryear{Veitch et al.}{2015}]{Veitch2015} Veitch J., Raymond V., Farr B., Farr W., Graff P., Vitale S., Aylott B., et al., 2015, PhRvD, 91, 042003. doi:10.1103/PhysRevD.91.042003

\bibitem[\protect\citeauthoryear{Vieira et al.}{2020}]{Vieira2020} Vieira N., Ruan J.~J., Haggard D., Drout M.~R., Nynka M.~C., Boyce H., Spekkens K., et al., 2020, ApJ, 895, 96. doi:10.3847/1538-4357/ab917d

\bibitem[\protect\citeauthoryear{Wang et al.}{2015}]{Wang2015} Wang X.-G., Zhang B., Liang E.-W., Gao H., Li L., Deng C.-M., Qin S.-M., et al., 2015, ApJS, 219, 9. doi:10.1088/0067-0049/219/1/9

\bibitem[\protect\citeauthoryear{Watson et al.}{2016}]{Watson2016} Watson A.~M., et al., 2016b, SPIE, 99100G

\bibitem[\protect\citeauthoryear{Watson et al.}{2019c}]{2019GCN.24086....1W} Watson A.~M., et al., 2019a, GCN, 24086, 1

\bibitem[\protect\citeauthoryear{Watson et al.}{2019b}]{2019GCN.24310....1W} Watson A.~M., et al., 2019b, GCN, 24310, 1

\bibitem[\protect\citeauthoryear{Watson, et al.}{2019a}]{2019GCN.24644....1W} Watson A.~M., et al., 2019c, GCN, 24644, 1

\bibitem[\protect\citeauthoryear{Watson et al.}{2020a}]{Watson2020} Watson A.~M., et al., 2020a, MNRAS, 492, 5916

\bibitem[\protect\citeauthoryear{Watson}{2020b}]{2020GCN.27061....1W} Watson A.~M., 2020b, GCN, 27061, 1

\bibitem[\protect\citeauthoryear{Wollaeger et al.}{2018}]{Wollaeger2018} Wollaeger R.~T., Korobkin O., Fontes C.~J., Rosswog S.~K., Even W.~P., Fryer C.~L., Sollerman J., et al., 2018, MNRAS, 478, 3298. doi:10.1093/mnras/sty1018

\bibitem[\protect\citeauthoryear{Wollaeger et al.}{2021}]{Wollaeger2021} Wollaeger R.~T., Fryer C.~L., Chase E.~A., Fontes C.~J., Ristic M., Hungerford A.~L., Korobkin O., et al., 2021, arXiv, arXiv:2105.11543

\bibitem[\protect\citeauthoryear{Yang et al.}{2019}]{Yang2019} Yang Y., Bartos I., Gayathri V., Ford K.~E.~S., Haiman Z., Klimenko S., Kocsis B., et al., 2019, PhRvL, 123, 181101. doi:10.1103/PhysRevLett.123.181101

\bibitem[\protect\citeauthoryear{Yi \& Cheng}{2019}]{Yi2019} Yi S.-X., Cheng K.~S., 2019, ApJL, 884, L12. doi:10.3847/2041-8213/ab459a


\bibitem[\protect\citeauthoryear{Yuan et al.}{2021}]{Yuan2021} Yuan Y., L{\"u} H.-J., Yuan H.-Y., Ma S.-B., Lei W.-H., Liang E.-W., 2021, arXiv, arXiv:2103.05811

\bibitem[\protect\citeauthoryear{Yu, Zhang, \& Gao}{2013}]{2013Yu} Yu Y.-W., Zhang B., Gao H., 2013, ApJL, 776, L40. doi:10.1088/2041-8205/776/2/L40

\bibitem[\protect\citeauthoryear{Zhang}{2016}]{Zhang2016} Zhang B., 2016, ApJL, 827, L31. doi:10.3847/2041-8205/827/2/L31

\bibitem[\protect\citeauthoryear{Zhang}{2019}]{Zhang2019} Zhang B., 2019, ApJL, 873, L9. doi:10.3847/2041-8213/ab0ae8

\bibitem[\protect\citeauthoryear{Zou et al.}{2018}]{Zou2018} Zou Y.-C., Wang F.-F., Moharana R., Liao B., Chen W., Wu Q., Lei W.-H., et al., 2018, ApJL, 852, L1. doi:10.3847/2041-8213/aaa123

\end{thebibliography}



\bsp	
\label{lastpage}
\end{document}